		\documentclass[10pt,final,doubleside]{IEEEtran}
	\usepackage{pifont,bm,multicol,amsfonts,amsmath,color,amssymb,graphicx, epsfig,cite,psfrag,subfigure,algorithm,enumerate,stfloats,algorithm,algorithmic,
epstopdf,balance}
\usepackage{tabularx}
	\usepackage{mathrsfs,pifont,bm,multicol,amsfonts,amsmath,color,amssymb,graphicx, epsfig,cite,psfrag,subfigure,algorithm,enumerate,stfloats,algorithm,algorithmic,
epstopdf,balance}

	\newtheorem{remark}{\underline{Remark}}

\begin{document}
\title{STAR-RIS Aided Dynamic Scatterers Tracking for Integrated Sensing and Communications
}
%{ Dynamic Scattering Environment Tracking and Communication Enhancement for STAR-RIS ISAC}
\author{

\thanks{
Manuscript received 26 June, 2024; revised 6 September, 2024, accepted 6 January, 2025; date of current version 7 January, 2024.
The work of Muye Li and Shun Zhang was supported in part by the National Key Research and Development Program of China under Grant 2023YFB2904500, and in part by the National Natural Science Foundation of China under Grant 62271373.
The work of Yao Ge was supported by the RIE2020 Industry Alignment Fund-Industry Collaboration Projects (IAF-ICP) Funding Initiative, as well as cash and in-kind contribution from the industry partner(s).
The work of Chau Yuen was supported by Ministry of Education, Singapore, under its MOE Tier 2 (Award number MOE-T2EP50220-0019) and National Research Foundation, Singapore and Infocomm Media Development Authority under its Future Communications Research $\&$ Development Programme FCP-NTU-RG-2024-025.
\emph{(Corresponding author: Shun Zhang.)}
%The work of P. Fan was supported by NSFC Project (No.62020106001).
%\emph{(Corresponding author: Shun Zhang.)}
%This work was supported in part by the National Natural Science Foundation of China under Grant 61871455, 61931017, 6183101, 61901329, in part by the China Postdoctoral Science Foundation under Grant 2019M653557, in part by NSFC Project (No.61731017) and National Key R$\&$D Project (No.2018YFB1801104), in part by the Fundamental Research Funds for the Central Universities and the Innovation Fund of Xidian University under Grant 20109205456, also in part by the SAIC Science and Technology Foundation under Grant No. 1911.
}

Muye Li, Shun Zhang, \emph{Senior Member, IEEE}, Yao Ge, \emph{Member, IEEE}, and Chau Yuen, \emph{Fellow, IEEE}

%Muye Li, Shun Zhang, \emph{Senior Member, IEEE}, Yao Ge, \emph{Member, IEEE}, Zan Li, \emph{Senior Member, IEEE}, Feifei Gao, \emph{Fellow, IEEE}, Pingzhi Fan, \emph{Fellow, IEEE}

\thanks{Muye Li and Shun Zhang are with the State Key Laboratory of Integrated Services Networks, Xidian University, Xi'an 710071, China (e-mail: myli$\_$96@stu.xidian.edu.cn; zhangshunsdu@xidian.edu.cn).}

\thanks{Yao Ge is with the Continental-NTU Corporate Laboratory, Nanyang Technological University, Singapore 637553 (e-mail: yao.ge@ntu.edu.sg).}

\thanks{Chau Yuen is with the School of Electrical and Electronics Engineering, Nanyang Technological University, Singapore 639798 (e-mail: chau.yuen@ntu.edu.sg).}

}

\maketitle
\vspace{-3mm}
\begin{abstract}
Integrated sensing and communication (ISAC) has become an attractive technology for future wireless networks.
In this paper, we propose a simultaneous transmission and reflection reconfigurable intelligent surface (STAR-RIS) aided dynamic scatterers tracking scheme for ISAC in high mobility millimeter wave communication systems, where the STAR-RIS is employed to provide communication service for indoor user with the base station (BS) and simultaneously sense and track the interested outdoor dynamic scatterers.
Specifically, we resort to an active STAR-RIS to respectively receive and further deal with the impinging signal from its double sides at the same time.
Then, we develop a transmission strategy with the activation scheme of the STAR-RIS elements, and construct the signal models within the system.
After acquiring the channel parameters related to the BS-RIS channel, the dynamic paths can be identified from all the scattering paths, and the dynamic targets can be classified with respect to their radar cross sections.
We further track the outdoor scatterers at STAR-RIS by resorting to the Gaussian mixture-probability hypothesis density filter.
With the tracked locations of the outdoor scatterers, a beam prediction strategy for both the precoder of BS and the refraction phase shift vector of STAR-RIS is developed to enhance the communication performance of the indoor user.
Besides, a target mismatch detection and path collision prediction mechanism is proposed to reduce the training overhead and improve the transmission performance.
Finally, the feasibility and effectiveness of our proposed STAR-RIS aided dynamic scatterers tracking scheme for ISAC are demonstrated and verified via simulation results.

\end{abstract}

\maketitle
\thispagestyle{empty}


\section{Introduction}
In the past few decades, two representative applications, i.e., wireless communications and radar sensing, have respectively achieved remarkable improvements.
With the rapid growth of the connected devices and the limitation of spectrum resources, integrated sensing and communication (ISAC) has become a hot spot research in recent years \cite{ISAC_Mag1,ISAC_Mag2,ISAC_Mag3,ISAC_Mag4,ISAC_Mag6}.
Compared with the traditional independent communication and sensing systems, the advantages of ISAC lies on two kinds of aspects.
On the one hand, since sensing and communication systems employ similar radio frequency (RF) front-end hardware and similar signal processing techniques \cite{ISAC_advan1}, ISAC can realize the sharing of the rare spectrum resource and the saving of hardware cost.
On the other hand, signal processing units in communication system can achieve sensing functionalities \cite{ISAC_advan21}, and the knowledge of environment sensing information can also significantly benefit communication performance \cite{ISAC_advan22}.
Incorporating with other state-of-art technologies, such as millimeter wave (mmWave) \cite{ISAC_integrated1}, Teraherz\cite{{ISAC_THz}}, non-orthogonal multiple access (NOMA) \cite{ISAC_integrated2}, and artificial intelligence \cite{ISAC_integrated3}, ISAC can achieve higher sensing accuracy and improve spectral efficiency.
Therefore, ISAC has become a promising candidate technology for future wireless networks.

To exploit the integration gains of ISAC systems, the inherent conflicting requirements of sensing and communication should be carefully balanced in order to achieve acceptable trade-offs \cite{ISAC_m_MIMO1}.
To address this issue, massive multi-input multi-output (MIMO) architectures have been employed to provide spatial degrees of freedom (DoFs) \cite{ISAC_m_MIMO2}, which makes it possible to simultaneously serve multiple targets and communication users.
In \cite{mMIMO_com_1}, Shahmansoori \emph{et al.} studied the performance bound of position and rotation angle estimation in the presence of scatterers, and proposed a matching pursuit based algorithm in a mmWave MIMO system.
%From a compressed sampling perspective, Gao \emph{et al.} proposed a ISAC processing framework in \cite{mMIMO_com_2}, where both the high dimensional channel state information and the radar imaging information can be simultaneously recovered with reduced pilot overhead.
In \cite{mMIMO_com_3}, Xu \emph{et al.} proposed a spatial structure matching based angle of arrival (AoA) estimation and tracking scheme, which can adapt to the dynamic changes of scattering environment.
Jointly considering the beam-squint and beam-split effect caused by large antenna array and wide bandwidth, Gao \emph{et al.} proposed an ISAC scheme with significantly reduced training cost \cite{mMIMO_com_4}.

However, in actual uncontrollable electromagnetic environments, since the electromagnetic waves are usually blocked by unexpected physical blockages, it is inevitable to experience performance deteriorations \cite{ISAC_RIS1}.
Fortunately, reconfigurable intelligent surface (RIS) can perfectly adapt and tackle such a challenge.
On the one hand, RIS can actively control the incident electromagnetic wave, adjust the propagation environment, and strengthen the desiring scattering paths \cite{ISAC_RIS_advan1}.
On the other hand, if two terminals of an ISAC link are completely blocked by physical obstacles, RIS can help maintain their link by artificially constructing virtual line-of-sight (LoS) links \cite{ISAC_RIS_advan2}.
Thus, RIS is very suitable for ISAC systems, and has been tightly combined with ISAC in recent studies.
In \cite{RIS_com_1}, Shao \emph{et al.} proposed a self-sensing RIS architecture, where the RIS can sense the direction of nearby target applying a customized multiple signal classification (MUSIC) algorithm with the equipped dedicated sensors.
To simultaneously improve the performance of communication and sensing, Xu \emph{et al.} proposed two kinds of alternating optimization (AO) based beamforming designs in RIS-assisted ISAC systems \cite{RIS_com_2}.
%In \cite{RIS_com_3}, Guo \emph{et al.} formulated a problem for joint design of beamforming, power allocation and signal processing in a full-duplex RIS-aided uplink communication system, and proposed an iterative solution with convex optimization techniques.
For the ISAC service of a blind-zone user, Qian \emph{et al.} proposed a RIS-aided ISAC framework in \cite{RIS_com_4}, where a two-phase transmission protocol for location sensing and beamforming design was developed to realize desired ISAC performance.

Nevertheless, there is a crucial limitation on the utilization of traditional RIS in ISAC, i.e., the sensing targets as well as the communication terminals should be at the same side of RIS \cite{ISAC_STARRIS_advan1}.
Thus, it cannot achieve full-space coverage.
To resolve this limitation, the simultaneously transmitting and reflecting RIS (STAR-RIS) is proposed, where communication terminals and sensing targets can be located at double sides of the STAR-RIS.
Therefore, STAR-RIS aided ISAC can provide more DoFs than that with traditional RIS.
%In \cite{STARRIS_com_1}, Wang \emph{et al.} used STAR-RIS to partition the entire space into a sensing space and a communication one, and derived the minimal Cramer-Rao bound of DoA estimation subject to the minimum communication requirement.
Focusing on uncontrollable electromagnetic environment, Liu \emph{et al.} considered an energy splitting STAR-RIS empowered ISAC network in \cite{STARRIS_com_2}, where the beamforming at base station (BS) and the phase shifts at STAR-RIS are jointly optimized for maximizing the sensing SNR.
In \cite{STARRIS_com_3}, Wang \emph{et al.} investigated the fairness of STAR-RIS and NOMA assisted ISAC systems, and proposed to eliminate the interference of sensing signal before decoding the communication signals.
Considering the security issues with multiple potential eavesdroppers,
Wang \emph{et al.} further proposed a symbol-level precoding scheme in \cite{STARRIS_com_4} for concurrent securing confidential information transmission and target sensing.

For high mobility scenario, the communication terminals as well as the sensing targets may move quickly, and the scattering environment will become time-variant in ISAC systems \cite{ISAC_mobility_charac1}.
On the one hand, one vehicle expects the knowledge about the position and velocity of other nearby vehicles in order to predict their trajectory and schedule the later motion of itself for future autonomous vehicles networks \cite{ISAC_mobility_fact1}.
On the other hand, the moving targets can also be part of the scatters in the surrounding communication environment,
and thus, the sensing and tracking of these targets can return to help the prediction of the communication channel and further enhancement of communication transmission \cite{ISAC_mobility_fact2}.
Focusing on vehicular networks, Liu \emph{et al.} proposed a general point-to-point ISAC model in \cite{mobility_com_1} to account for the scenarios where sensing state is different from
the correlated channel state.
In \cite{mobility_com_4}, Li \emph{et al.} proposed to sense the location and velocity of a moving vehicle aided by a STAR-RIS, and designed a trade-off scheme for the performance of sensing and communication.
In \cite{mobility_com_2}, Ronquillo \emph{et al.} proposed to actively sense and sequentially track the beams between a quasi-stationary receiver and mobile transmitters at mmWave frequencies and above.
Considering the geometry of the vehicles, Du \emph{et al.} proposed to cover entire vehicle in real-time with the designed beams to improve communication performance in \cite{mobility_com_3}.
%In \cite{mobility_com_4}, Li \emph{et al.} considered RIS-aided downlink ISAC, and proposed a hybrid BS beamforming and RIS phase shift optimization problem to improve the system sum rate with hardware and sensing quality of service limitations.

%\subsection{Related Works}

\subsection{Motivations and Contributions}

The aforementioned works only focused on the mobility of the sensing targets.
However, due to the limitation of the ISAC serving area for a particular BS, the moving targets tend to frequently enter or exit the area.
On the other hand, existing targets may also merge/spawn a number of targets, such as passengers getting on/off a bus.
Therefore, the variation and tracking of the targets from one sensing period to the next should be both carefully dealt with, including the birth and death of targets.
Besides, traditional sensing functionalities are always implemented at the BS or radar by utilizing the received echo signal.
However, although the echo signal contains the information about the targets that constitute communication channel, clutters and the reflected signal from some uninterested targets are also included.
For high ISAC efficiency, the ISAC system should only focus on the targets that would influence the communication channel.
To address this challenge, one effective approach is to integrate sensing capabilities into a communication node within the system, such as the STAR-RIS between the BS and its serving communication users.

Based on the above mentioned motivation,
our work focuses on the STAR-RIS aided downlink ISAC scenario, and proposes an efficient tracking scheme for a variation number of dynamic targets, which can help to further enhance the efficiency and overall performance of ISAC systems.
In particular, the sensing functionality is implemented at an active STAR-RIS to mitigate the influence of undesired targets that do not contribute to the communication links.
Main contributions of our work are outlined as follows:
\begin{itemize}
  \item
  We propose a STAR-RIS aided ISAC framework over mobility scenario, where the STAR-RIS is equipped with a small number of radio-frequency (RF) chains, and is utilized to simultaneously enhance the communication performance of the indoor user and sense the outdoor dynamic targets.
%  Thus, the STAR-RIS is able to absorb and deal with the impinging signal from its double sides at the same time, respectively.
%  The STAR-RIS is deployed at the surface of a building, dividing the entire space into the outdoor space with a varying number of scatterers (also known as the targets to be tracked) and the indoor space with a user terminal (UT).
  Based on the system configuration, we model the movement of the scatterers as state transition equations, and propose an efficient transmission frame structure and a specific STAR-RIS elements activation strategy to track the dynamic scatterers.

  \item
  By utilizing the knowledge of angular information, i.e., the angle of departures (AoDs) of the BS-RIS link, we first propose to identify the dynamic paths among all the scattering paths, and then classify the dynamic scatterers based on their radar cross sections (RCSs).
  Furthermore, we track the scatterers (including the targets' amount and their corresponding locations) at the STAR-RIS by utilizing the Gaussian mixture-probability hypothesis density (GM-PHD) filter.

  \item
  To further enhance communication performance,
  we propose a beam prediction strategy for both the precoder of BS and the refraction phase shift vector of STAR-RIS by taking advantage of the sensing results for the outdoor scatterers.
%  On the other hand, the second strategy is to predict the overall channel with the sensing results, and then optimize the predicted beam based on this.
  Furthermore, we develop a target mismatch detection and path collision prediction mechanism to reduce the training overhead and improve transmission performance.

  \item
  Simulation results are provided to verify the effectiveness of our proposed STAR-RIS aided dynamic scatterers tracking scheme for ISAC.
  Our results also demonstrate that the proposed beam tracking scheme can achieve better communication performance than that without beam tacking scheme (i.e., fixed beamforming).

\end{itemize}

\subsection{Organizations and Notations}

The rest of this paper is organized as follows.
Section II illustrates the STAR-RIS aided downlink ISAC system model over mobility scenario.
Then, the multiple target tracking scheme is proposed by using GM-PHD filter in Section III.
Section IV introduces the proposed beamforming and phase shifts prediction strategies of the BS and the STAR-RIS, and illustrates the proposed mismatch detection and path collision prediction mechanism.
In Section V, simulation results are provided and discussed.
Finally, conclusions are drawn in Section VI.

Notations: Denote lowercase (uppercase) boldface as vector (matrix).
$(\cdot )^H $, $(\cdot )^T $, $(\cdot )^{*} $, and $(\cdot )^{\dagger} $ represent the Hermitian transform, transpose, conjugate, and pseudo-inverse, respectively.
$\mathbf I_N $ is an $N \times N $ identity matrix.
$\mathbb E \{\cdot \} $ is the expectation operator.
Denote $|\cdot | $ as the amplitude of a complex value.
$[\mathbf A]_{i,j} $ and $\mathbf A_{\mathcal Q,:}$ (or $\mathbf A_{:, \mathcal Q} $) represent the $(i,j) $-th entry of $\mathbf A $ and the submatrix of $\mathbf A $ which contains the rows (or columns) with the index set $\mathcal Q $, respectively.
$\mathbf x_{\mathcal Q} $ is the subvector of $\mathbf x $ built by the index set $\mathcal Q $.
%$\lfloor p \rfloor $ denotes the largest integer less than or equal to $p $.
%${\boldsymbol\Xi}^{(l-1)} \setminus {\boldsymbol \alpha}^{(l-1)}$ denotes the set ${\boldsymbol\Xi}^{(l-1)}$ expect the element ${\boldsymbol \alpha}^{(l-1)}$.
%The real component of $x $ is expressed as $\Re \{x\}$.
$\text{diag} (\mathbf x)$ is a diagonal matrix whose diagonal elements are formed with the elements of $\mathbf x $.
$[\mathbf x]_i$ is the $i$-th entry of $\mathbf x$.
$(x)_n$ is the mod operation of $x$ with respect to $n$.

\section{STAR-RIS Aided ISAC System Model over Mobility Scenario}

\subsection{Scenario Configuration}

\begin{figure}[htbp]
 \centering
 \includegraphics[width=80mm]{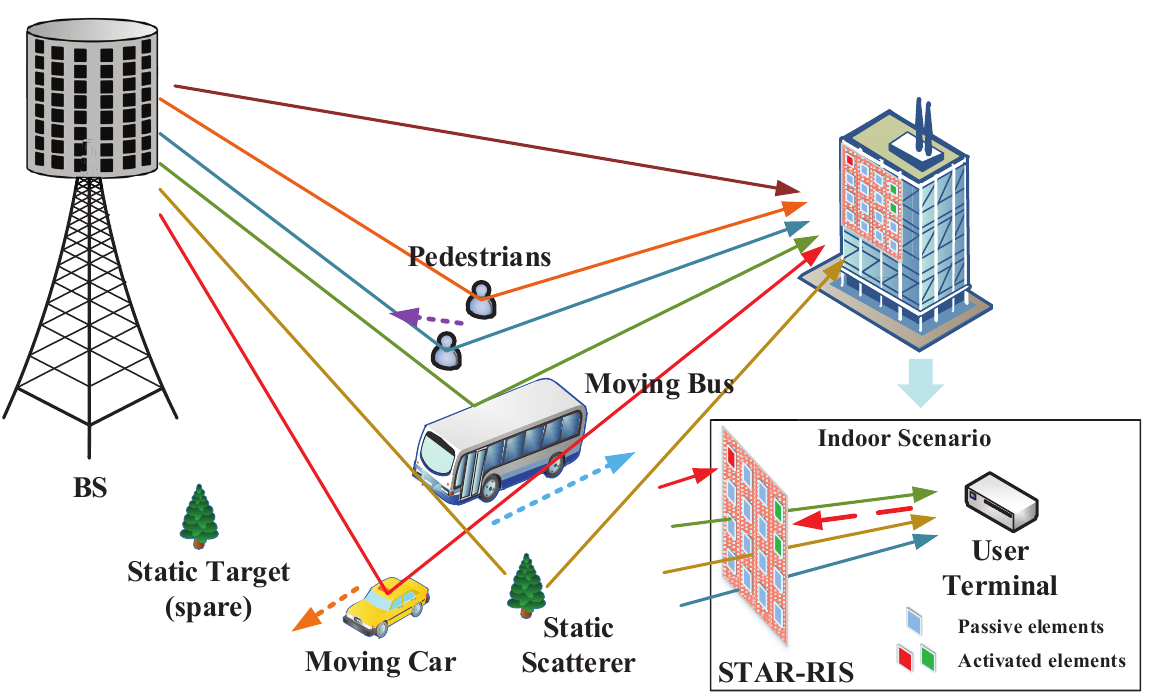}
 \caption{STAR-RIS aided indoor-outdoor ISAC system scenario.}
 \label{scenario}
\end{figure}

As shown in \figurename{ \ref{scenario}}, we consider a downlink mmWave communication system with a BS, a STAR-RIS, and a single-antenna indoor user terminal (UT),
where the STAR-RIS is utilized not only for enhancing the communication performance between the BS and the indoor UT, but also for sensing and tracking the outdoor dynamic scatterers.
Thus, the STAR-RIS is deployed at the surface of a building, which separate the whole space into the indoor and outdoor parts, respectively.
The BS is equipped with $N_B = N_B^x \times N_B^y$ uniform plain array (UPA), and $N_B^{RF}$ radio frequency (RF) chains.
In the meantime, the STAR-RIS is equipped with $N_R = N_R^x \times N_R^y$ UPA and $N_S^{RF}$ RF chains.
Note that $N_S^{RF}-1$ RF chains are utilized to alternatively activate partial STAR-RIS elements for the communication of the indoor UT, while the remained one RF chain is to activate a fixed element for the sensing of outdoor scatterers.
%where one of the RF chains is to activate a fixed element for receiving outdoor signal, and the others are to alternatively activate partial STAR-RIS elements for the indoor signal.
Compared to sensing at the BS with echo signal, sensing at the STAR-RIS can assure that all the detected targets contribute to the BS-RIS communication channel.
Such sensing information can be further used to improve communication performance.
Without loss of generality, it is assumed that there are at most $P_{max}$ scatterers in the communication environment.
Note that the number of scatterers at the $k$-th ISAC period is $P(k) \leq P_{max}$, where the velocity of the $p$-th scatterer is denoted by $\mathbf v_{k,p}$.
Some of the scatterers are dynamic, such as pedestrians and vehicles, while others are static, such as trees and buildings.
Besides, considering the actions of dynamic scatterers, e.g., pedestrians getting on/off the vehicles, entering or leaving the LoS region of the BS, the actual number of scatterers is also dynamic.
Hence, $P(k)$ may not be equal to $P(k-1)$.

\begin{figure}[htbp]
 \centering
 \includegraphics[width=80mm]{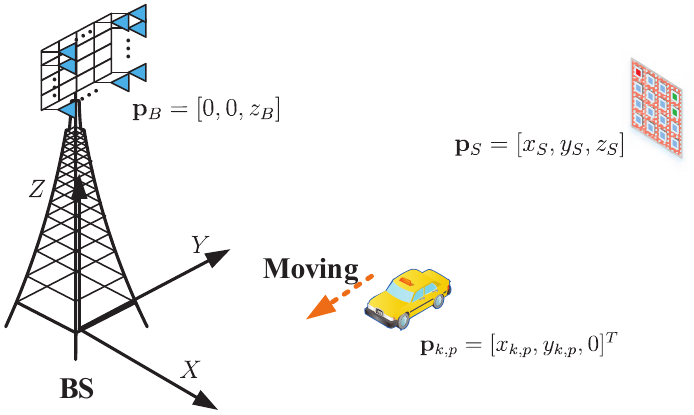}
 \caption{Illustration of the global Cartesian coordinate system.}
 \label{coordinate_system}
\end{figure}

For the notational convenience, we establish a global Cartesian coordinate system centered at the BS antenna plane, as shown in \figurename{ \ref{coordinate_system}}, whose $Z$-axis is perpendicular to the ground, and $YOZ$-plain is parallel to the UPA plane of the BS.
Accordingly, the coordinates of the BS and the STAR-RIS can be respectively represented as $\mathbf p_B = [0,0,z_B]^T$ and $\mathbf p_S = [x_S,y_S,z_S]^T$.
Moreover, the location of $p$-th scatterer at ISAC period $k$ is defined as $\mathbf p_{k,p} = [x_{k,p}, y_{k,p}, 0]^T$, and its state is represented by $\mathbf c_{k,p} = [x_{k,p}, y_{k,p}]^T$.

In real mobility scenario, the velocities of all the dynamic targets may vary between different sensing periods.
However, since the sensing period is always very short, it is reasonable to assume that the targets undergo a uniform linear motion with small fluctuations.
Take a general transportation scenario as an example.
Assume that the velocity and the acceleration of the target are $v_{k,p} = 10$ m/s and $a_{k,p} = 0.5$ m/s$^2$, respectively, and set the sensing period as $0.5$ s.
Thus, the actual movement distance of the target in one period would be $10 \ \text{m/s} \times 0.5 \ \text{s}+1/2 \times (0.5 \  \text{m/s})^2 \times (0.5 \  \text{s})^2 = 5.0625$ m, which is only $0.0625$ m different from that with uniform linear motion.
\footnote{Note that we only take the sensing period of 0.5s as an example in this manuscript.
However, it can be directly extended to scenarios with other reasonable settings.}
Hence, the state $\mathbf c_{k,p}$ can be regarded as a Gaussian linear model, i.e.,
\begin{align}\label{AR_model}
\mathbf c_{k|k-1,p} =& \mathbf A \mathbf c_{k-1,p} + \mathbf q_{k-1,p},
%\\
%\mathbf z_{p,k} =& \mathbf g(\mathbf c_{p,k|k-1}) + \mathbf r_{k}
\end{align}
where $\mathbf A\in \mathbb R^{2\times2}$ is the state transition matrix, $\mathbf q_{k-1,p} \!\in \mathbb R^{2\times1} \sim \mathcal N (\mathbf 0, \mathbf Q_{p})$ is the process noise, and
$\mathbf Q_{p} \in \mathbb R^{2\times2}$ is the covariance matrix of $\mathbf q_{k-1,p}$.
Thus, $\mathbf c_{k|k-1,p} \sim \mathcal N (\mathbf A \mathbf c_{k-1,p}, \mathbf Q_{p})$.
Especially, since we are considering the transition of target location, $\mathbf A$ and $\mathbf Q_{p}$ in \eqref{AR_model} can be respectively represented as
\begin{align}
\mathbf A = \mathbf I_2
\quad\quad\text{and}\quad\quad
\mathbf Q_{p} = \sigma_{v,p} \Delta \mathbf I_2,
\end{align}
where $\mathbf I_2$ denotes the $2\times 2$ identity matrix,
$\Delta$ is ISAC period, and $\sigma_{v,p}$ is the standard deviation of the process noise.

%Specially, we do not regard all the scatterers as particles at their centroid.
%In other word, the relatively small scatterers, i.e., pedestrians, are seen as centroids.
%However, relatively larger scatterers, i.e., the cars and buses, are modelled as the set of a number of scattering anchors, as shown in \textcolor[rgb]{1.00,0.00,0.00}{Figxxx}.
%Besides, considering the actions of dynamic scatterers, i.e., pedestrians getting on and off the vehicles, entering or leaving the line-of-sight (LOS) region of the BS, the actual number of scatterers is also dynamic.
%Thus, the tracking and prediction of the dynamic scatterers in the environment is important for communication.

\subsection{ISAC Transmission Frame Structure}

\begin{figure}[htbp]
 \centering
 \includegraphics[width=90mm]{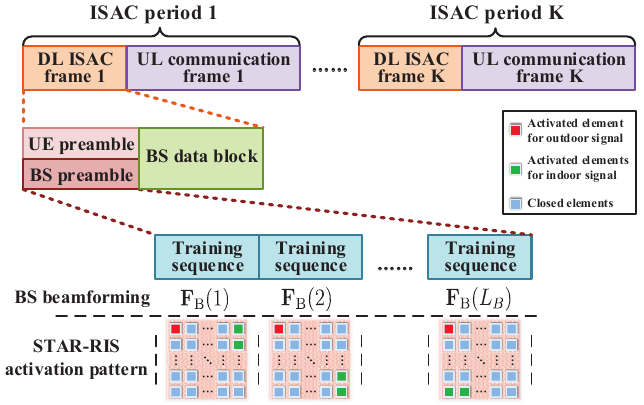}
 \caption{Proposed ISAC transmission structure.}
 \label{DL_trans_struc}
\end{figure}

The downlink signals transmitted by the BS will first be reflected by the scatterers, then reconfigured by the STAR-RIS, and finally received by the indoor UT.
To track the dynamic scatterers and further improve communication performance, we develop an efficient ISAC transmission structure, as depicted in \figurename{ \ref{DL_trans_struc}}.
Each ISAC period consists of a downlink (DL) ISAC frame and an uplink (UL) communication frame.
We assume that the scattering environment does not change during each ISAC period.
In each DL frame, BS data symbols are transmitted after the preamble.
Within the preamble duration, the BS and UT will both transmit multiple training sequences to the STAR-RIS, where the sequences transmitted by the BS are precoded by orthogonal precoders $\mathbf F_B(l_B)$ with different sequence index $l_B$.
Denote the size of beam scanning codebook for the BS as $N_B$, it can be checked that the maximal required number of training sequences is $L_B = N_B / N_B^{RF}$.
On the other hand, the settings for the preamble of UT are the same with that of the BS.

In the preamble phase, the STAR-RIS will respectively receive the impinging signal from both the indoor UT and the outdoor BS via different activated elements.
Note that the inactive elements will be closed at that time, and will not let any signal penetrate the STAR-RIS.
As for the training sequences transmitted by the BS, the STAR-RIS will activate a fixed element to receive the outdoor signal and detect the outdoor signal strength, determine the outdoor scattering paths, and estimate the outdoor channel parameters, i.e., the Doppler frequency shifts and the delays, for further sensing and tracking.
On the other hand, the STAR-RIS will activate $N_S^{RF}\!-\!1$ elements for one training sequence transmitted by the indoor UT.
Note that different elements will be activated for different training sequences, and one row and one column elements will be alternatively activated during the whole preamble to reconstruct the complete indoor channel, similar to \cite{ISAC_RIS_advan2}.
Thus, the number of training sequences transmitted by the UT is $L_U = \lceil(N_R^x + N_R^y-1) / (N_S^{RF} - 1)\rceil$.
In the data transmission phase, the BS will implement beamforming towards the detected scattering paths.
In the meantime, the STAR-RIS will be in full passive mode, and focus the impinging communication signal to the indoor UT by adjusting the refraction phase shift vector.

\begin{remark}
Note that the number of training sequences can be reduced in practice for subsequent ISAC periods.
For a specific ISAC period, the outdoor scatterers have been already tracked and predicted in the last period.
Thus, to keep the tracking performance and simultaneously reduce the training overhead,
we can only scan the beam directions around those scatterers and ignore the searching of other directions.
For description simplicity, we still apply the full-space beam scanning scheme in the transmission structure.
\end{remark}
%\textcolor[rgb]{1.00,0.00,0.00}{During the preamble, the HRIS is in ctive mode, where $N_R^{RF}$ elements are activated.
%In the meantime, other elements are still in passive mode and can reflect the impinging electromagnetic wave to the UT.
%with the received signal, both the HRIS and the UT will detect the signal strength, determine the scattering paths, and estimate the channel parameters, i.e., the Doppler frequency shifts and the delays.
%In the data transmission phase, the BS will implement beamforming towards the detected scattering paths.
%The HRIS will be in fully passive mode and focus the impinging signal to the indoor UT.
%And the UT will demodulate the receiving signals for data detection.}

\subsection{Channel Model}

Note that the channel model is general for different scatterer amounts.
Since the number of scatterers is varying with respect to the ISAC period index $k$, we focus on a specific ISAC period and omit $k$ in the derivations for notational simplicity.
The channel between the BS and the UT is cascaded by the STAR-RIS.
Due to the dynamic scattering environment, the channel from the BS to the STAR-RIS is time-varying, thus its corresponding time-frequency selective fading outdoor channel can be expressed as
\begin{align}\label{out_channel}
\mathbf h_i^{O}\!(t) \!=&\!\! \sum_{p=1}^{P} \!\!h_p^{\text{O}} e^{\jmath 2\pi \nu_p^{\text{O}} (t-\tau_p^{O})} \delta(iT_s \!\!- \!\!\tau_p^{O}\!) \mathbf a_O \!(\phi_p^S, \psi_p^S) \mathbf a_B^T \!(\phi_p^B, \psi_p^B\!),
\notag\\
&\quad\quad\quad\quad\quad\quad\quad\quad\quad\quad\quad\quad i = 0, \ldots, I-1,
\end{align}
where $i$ denotes the index of the delay tap, and $I$ is the maximum number of delay taps between the BS and the STAR-RIS.
$P$ is the number of scattering paths from the BS to the STAR-RIS.
$T_s$ is the system sampling period, and $\delta(\cdot)$ is the Delta function.
$h_p^{O} = \alpha_p e^{\jmath \beta_p}$ is the complex channel gain with $\alpha _p = \frac{\sigma_p \lambda^2}{(c\tau_p^O)^2}$ and $\beta_p$ being the amplitude and the phase of $h_p^{O}$, respectively, where $\sigma_p$, $\lambda$, and $c$ are the radar cross section (RCS), the signal wavelength, and the light speed, respectively.
%$h_p^{O} \sim \mathcal {CN} (0, \lambda_p^{\text{O}})$ is the complex channel gain with its variance $\lambda_p^{\text{O}}$,
$\tau_p^{O}$ and $\nu_p^{\text{O}}$ are the delay and Doppler frequency shift of the $p$-th path, respectively.
In addition, $\{\phi_p^B, \psi_p^B\}$ and $\{\phi_p^S, \psi_p^S\}$ are the elevation and azimuth AODs at the BS and those AOAs at the STAR-RIS, respectively.
The array steering vectors $\mathbf a_B(\phi, \psi) \in \mathbb C^{N_x^B N_y^B \times 1}$ and $\mathbf a_O(\phi, \psi) \in \mathbb C^{N_x^S N_y^S \times 1}$ are given by the form
$\mathbf a(\phi, \psi) = \mathbf a_x(\phi, \psi) \otimes \mathbf a_y(\phi, \psi) $, where $\otimes $ denotes the Kronecker product.
In addition, $\mathbf a_x(\phi, \psi)$ and $\mathbf a_y(\phi, \psi)$ are defined as
\begin{align}
  \mathbf a_x\!(\phi, \psi) \!\!=&\! [1, e^{\jmath 2\pi \frac{d}{\lambda} \!\sin(\phi) \!\sin(\psi)}, \ldots, e^{\jmath 2\pi(N_x\!-1\!) \frac{d}{\lambda} \!\sin(\phi) \sin(\psi)}]^T,
  \notag \\
  \mathbf a_y\!(\phi, \psi) \!\!=&\! [1, e^{\jmath 2\pi \frac{d}{\lambda} \!\sin(\phi) \!\cos(\psi)}, \ldots, e^{\jmath 2\pi(N_y\!-1\!) \frac{d}{\lambda} \!\sin(\phi) \cos(\psi)}]^T,
  \notag
\end{align}
where $d=\frac{\lambda}{2}$ is the inter-element spacing.
Note that the first element within the STAR-RIS UPA keeps being activated for the outdoor impinging signal during the preamble phase.

On the other hand, the indoor channel between the STAR-RIS and UT is almost static.
It is assumed that the LoS path is dominant.
Thus, the indoor channel can be represented as
\begin{align}\label{in_channel}
  \mathbf h_i^{\text{I}} =& h^{\text{I}} \delta(iT_s - \tau^{I}) \mathbf a_R^T(\phi^S, \psi^S),
  &i = 0, \ldots, I-1,
\end{align}
where $h^{I} \sim \mathcal {CN} (0, \lambda^{I})$ denotes the complex channel gain with the variance $\lambda^{I}$. $\tau^{I}$ is the delay of the LoS path, $\phi^S$ and $\psi^S$ are the elevation and azimuth AODs at the STAR-RIS, and $\mathbf a_R(\phi, \psi)$ is defined similar to $\mathbf a_O(\phi, \psi)$.

\subsection{Received Signal Model}
We analyze the received signal within one specific frame in this subsection.
Define the training sequence transmitted by the BS as $\mathbf t = [ t_0,  t_1, \ldots, t_{N_T-1}]^T \in \mathbb C^{N_T \times 1}$.
Then, the received signal at the STAR-RIS active element for the $n_T$-th time slot of the $l_B$-th training sequence within the preamble of one specific frame can be represented as
\begin{align}\label{HRIS_receive_nTnB}
{y}_{l_B}^{S}\!(n_T\!)\!\!=\!\! & \sum_{p=1}^P h_p^{{O}} e^{\jmath 2\pi \nu_p^{O} (((l_B-1) N_T+n_T) T_s-\tau_p^{O})} [\mathbf{a}_O(\phi_p^S, \psi_p^S)]_1
\notag \\
& \times \! \mathbf a_B^T(\phi_p^B, \psi_p^B)
 \mathbf F_B(l_B) \mathbf 1_{N_B^{RF}}
 [\mathbf{t}]_{n_T-\tau_p^{O} / T_s}\!\!+\!\!{w}_{l_B, n_T}^{S} \notag\\
=\!\! & \sum_{p=1}^P \!\overline{h}_p^{O} \!e^{\jmath 2\pi \nu_p^{O}((l_B-1) N_T+n_T) T_s} t_{(n_T-\tau_p^{O} / T_s)_{N_T}}\notag \\
& \!\times\!\! [\mathbf{a}_O \!(\!\phi_p^S, \!\psi_p^S)]_1 \mathbf a_B^T(\!\phi_p^B,\! \psi_p^B\!)
 \mathbf F\!_B(l_B \!) \mathbf 1\!_{N\!_B^{RF}}
\!\!+\!\! {w}_{l_B, n_T}^{S},
\end{align}
where $n_T = 0,1,..., N_T-1$, and $\overline{h}_p^{O} = h_p^{{O}} e^{\jmath 2\pi \nu_p^{O} \tau_p^{O}}$ is the equivalent channel complex gain.
$\mathbf 1_{N_B^{RF}} \in \mathbb N^{N_B^{RF} \times 1}$ is an all one vector, representing the baseband digital precoder of the BS.
Besides, $\mathbf F_B(l_B ) \in \mathbb C^{N_B\times N_B^{RF}}$ is the analog precoder of BS.
The subscript $(\cdot)_{N_T}$ of $t$ denotes the circular shift with respect to $N_T$.
Besides, ${w}_{l_B, n_T}^{S}$ is the additive Gaussian white noise (AWGN) with zero mean and variance $\sigma_n^2$.
To guarantee the estimation for the channel gain of any possible delay taps, it is assured that $N_T \geq \tau_{max}^{O}/T_s$, where ${\tau_{max}^{O}}$ is the maximum possible channel delay between the BS and the STAR-RIS.

Define the Doppler phase shift vector with respect to the preamble as $\mathbf{v}_T(\nu_p^{O}) = [1, e^{\jmath 2 \pi \nu_p^{O} T_s}, \ldots, e^{\jmath 2 \pi \nu_p^{O}(N_T-1) T_s}]^T$,
and reorganize the training vector for $\tau_p^{O}$ as $\mathbf{t}_{\tau} (\tau_p^{O}) = [t_{(0-\tau_p^{O} / T_s)_{N_T}}, \ldots, t_{(N_T-1-\tau_p^{O} / T_s)_{N_T}}]^T$.
Besides, define $f_{l_B} = (l_B -1) N_T$,
then the received outdoor signal at the STAR-RIS active element during the $l_B$-th training sequence can be derived by gathering \eqref{HRIS_receive_nTnB} with $n_T = 0, 1, \ldots, N_T-1$, which can be written as
\begin{align}\label{HRIS_receive_nB}
\mathbf{y}_{l_B}^{S}
= & \sum_{p=1}^P \bar{h}_p^{O} e^{\jmath 2\pi \nu_p^{O}f_{l_B} T_s} \left(\mathbf v_T(\nu_p^{O}) \odot \mathbf t_{\tau}(\tau_p^{O}) \right) \notag \\
& \times \!\![\mathbf{a}_O(\phi_p^S, \psi_p^S)]_1 \mathbf a_B^T(\phi_p^B, \psi_p^B)
 \mathbf F_B(l_B) \mathbf 1_{N_B^{RF}} \!+\! \mathbf{w}_{l_B}^{S},
\end{align}
where $\mathbf{w}_{l_B}^{S}\in \mathbb C^{N_T\times 1}$ is the AWGN vector of the $l_B$-th training sequence with zero mean and variance $\sigma_n^2$.

Moreover, define the data sequence as $\mathbf d = [d_0, d_1, \ldots, d_{N_D -1}]^T \in \mathbb C^{N_D \times 1}$, and define its reorganized form as $\mathbf{d}_{\tau} (\tau_p^{O}) = [d_{(0-\tau_p^{O} / T_s)_{N_D}}, \ldots, d_{(N_D-1-\tau_p^{O} / T_s)_{N_D}}]^T$.
In the mean time, the Doppler phase shift vector with respect to the data sequence is defined as $\mathbf{v}_D(\nu_p^{O}) = [1, e^{\jmath 2 \pi \nu_p^{O} T_s}, \ldots, e^{\jmath 2 \pi \nu_p^{O}(N_D-1) T_s}]^T$.
According to \eqref{in_channel} and \eqref{HRIS_receive_nB}, the received data signal at the UT within the data sequence can be represented as
\begin{align}\label{UT_receive_nB}
\mathbf{y}^{U}
= & h^{I}\sum_{p=1}^P \bar{h}_p^{O} \left(\mathbf v_D(\nu_p^{O}) \odot \mathbf d_{\tau}(\tau_p^{O}+\tau^{I}) \right) \notag \\
& \times
\mathbf a_R^T(\phi^S, \psi^S)
\text{diag}({\boldsymbol \omega})
\mathbf{a}_O(\phi_p^S, \psi_p^S) \mathbf a_B^T(\phi_p^B, \psi_p^B)
\notag\\
&\times {\mathbf F}_B \mathbf 1_{N_B^{RF}} + \mathbf{w}^{U},
\end{align}
where $\boldsymbol \omega \in \mathbb C^{N_R \times 1}$ is the STAR-RIS phase shift vector during the data transmission phase, and $\mathbf{w}^{U} \in \mathbb{C}^{N_D \times 1}$ is the AWGN vector at the UT with zero mean and variance $\sigma_n^2$.

With the received signal of all $L_B$ training sequences at the STAR-RIS, the outdoor channel parameters, i.e., $\boldsymbol \Xi_{p,k} = \{\phi_p^B, \psi_p^B, \tau_p^{O}, \nu_p^{O}, h_p^O\}, p = 1,2,\ldots, P(k)$ for each frame, can be estimated for sensing process and be used to further improve data transmission.
In this work, we focus on the realization of sensing and the enhancement of communication.
For more details about the channel parameter estimation, one can refer to \cite{mobility_com_4}.

% %可在数据阶段也用HRIS进行参数/强度的跟踪？
%Similarly, the received data signal at the UT in the same frame can be represented as
%\begin{align}\label{UT_receive_data_nB}
%\mathbf{y}_{d}^{U}
%= & h^{RU}\sum_{p=1}^P \bar{h}_p^{BR} e^{\jmath 2\pi \nu_p^{BR}f){n_B} T_s} \left(\mathbf v(\nu_p^{BR}) \odot \mathbf t_t(\tau_p^{BR}+\tau^{RU}) \right) \notag \\
%& \times
%[\mathbf a_R^T(\phi^R(t), \psi^R(t))]_{\overline{\mathbf b}}
%\text{diag}(\boldsymbol \omega_{\overline{\mathbf b}})
%[\mathbf{a}_R(\phi_p^R, \psi_p^R)]_{\overline{\mathbf b}} \mathbf a_B^T(\phi_p^B, \psi_p^B)
% \mathbf f_B(n_B) + \mathbf{w}_{n_B}^{U},
%\end{align}

\section{Gaussian Mixture-Probability Hypothesis Density Filter for Multi-Target Tracking}
\begin{figure*}[htbp!]
 \centering
 \includegraphics[width=140mm]{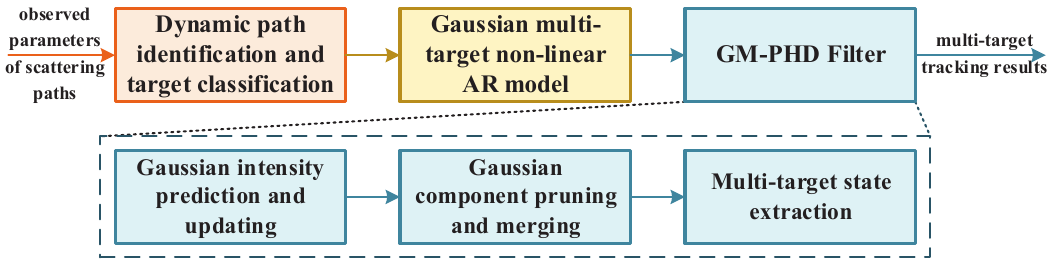}
 \caption{Procedure of the GM-PHD for multi-target tracking scheme.}
 \label{flowing_fig_GMPHD}
\end{figure*}

Assume that the outdoor channel parameters of each frame can be accurately estimated with the preambles.
As time goes by, the dynamic targets will move from one location to another, and it is necessary to track the dynamic scattering paths, predict the channel state information in subsequent ISAC periods, and maintain the communication performance.
On the other hand, the static scattering paths will not change in a long duration.
Thus, with the knowledge of outdoor channel parameters $\boldsymbol \Xi_{p,k}$, we should first distinguish the dynamic scattering paths, and then track the moving targets.
Although classical Kalman filter \cite{KF} can be resorted for tracking problem, it can not deal with the tracking of time varying number of targets.
Besides, although particle filtering can deal with the tracking varying number of targets \cite{particle_filtering}, the generation of a large number of particles will cause tremendous overhead.
Thus, we turn to the principle of GM-PHD filter for the tracking of multiple dynamic targets with varying amount.
The overall procedure of the GM-PHD filter is illustrated in \figurename{ \ref{flowing_fig_GMPHD}}.

%\begin{figure}[htbp]
% \centering
% \includegraphics[width=85mm]{flowing_fig.eps}
% \caption{Overall Procedure of the multi-target tracking scheme. \textcolor[rgb]{1.00,0.00,0.00}{\{\bf need revise\}}}
% \label{flowing_fig}
%\end{figure}

\subsection{Dynamic Paths Identification and Target Classification}

Since we only focus on dynamic targets, we should first separate the dynamic scattering paths from the obtained channel parameters, which can further reduce the computational complexity of target tracking.
It can be checked that the paths with zero Doppler frequency shift are related to static scatterers, while those with non-zero Doppler frequency shifts are relative to dynamic targets.
Thus, we only focus on the measurement of the scattering paths with non-zero Doppler frequency shifts.

Then, we focus on the target classification.
Since the BS and the STAR-RIS are usually deployed very high comparing to the sensing targets, i.e., the vehicles and pedestrians, it can be assumed that the RCS of one target is determined by its upper surface area.
Moreover, since there is a significant difference between the upper surface area of different targets, we can classify the targets as tiny targets (pedestrians), middle target (cars), and big target (buses) for further use by analyzing their measured RCSs.
This classification can be implemented by clustering methods \cite{cluster}.
After the classification, we will apply the auto-regressive (AR) model for multiple target in the next subsection.

%Define the velocity of the $k$-th target as $\mathbf v_k$.
%
%\textcolor[rgb]{1.00,0.00,0.00}{the channel gain is related with the signal travelling distance, and the RCS.
%If several beams with neighbored AODs have similar delays(distance) and strength, it can be checked that the RCS related with those beams are similar, and thus}
%
%
%
%judge the resolution of one beam

\subsection{Gaussian Multi-target Non-linear AR Model}

The dynamic characteristic of each moving target has been modelled as \eqref{AR_model} in Sec. II.
Besides, the measurement $\mathbf z_{k,p} = [\phi_{p}^B, \psi_{p}^B]^T$ follows a nonlinear Gaussian measurement model, i.e.,
\begin{align}\label{AR_model_meas}
\mathbf z_{k,p} =& \mathbf g(\mathbf c_{k|k-1,p}) + \mathbf r_{k,p},
\end{align}
where $\mathbf r_{k,p} \sim \mathcal N (\mathbf 0, \mathbf R_{k,p})$ is the measurement noise, and $\mathbf R_{k,p}$ is the covariance matrix of $\mathbf r_{k,p}$.
Hence, it can be checked that $\mathbf z_{k,p} \sim \mathcal N (\mathbf g(\mathbf c_{k|k-1,p}), \mathbf R_{k,p})$.
In addition, define $\mathcal Z_k$ as the measurement set in the $k$-th ISAC period for further use.
Besides, $\mathbf g(\cdot)$ is a nonlinear measurement function.
According to the geometric relationships between the locations and the measured angles, the entries of $\mathbf g(\cdot)$ can be respectively expressed as
\begin{align}
&[\mathbf g(\mathbf c_{k|k-1,p})]_1
\notag\\
&= \arccos
\frac{((\mathbf p_{k,p} - \mathbf p_{B}) - \frac{(\mathbf p_{k,p} - \mathbf p_{B})^T\mathbf e_Y}{\|\mathbf e_Y\|^2} \mathbf e_Y)^T \mathbf e_X}
{\|(\mathbf p_{k,p} - \mathbf p_{B}) - \frac{(\mathbf p_{k,p} - \mathbf p_{B})^T\mathbf e_Y}{\|\mathbf e_Y\|^2}\mathbf e_Y\| \cdot \|\mathbf e_X\|},
\label{measure_fun1}
\\
&[\mathbf g(\mathbf c_{k|k-1,p})]_2 \!=\! \frac{\pi}{2} -\! \arccos \frac{(\mathbf p_{k,p} - \mathbf p_{B})^T \mathbf e_Y}{\|\mathbf p_{k,p} - \mathbf p_{B}\| \cdot \|\mathbf e_Y\|},
\label{measure_fun2}
%\\
%[\mathbf g(\mathbf c_{k|k-1,p(k)})]_3 =& \frac{1}{c} (\|\mathbf p_{k,p(k)} - \mathbf p_B\| + \|\mathbf p_{k,p(k)} - \mathbf p_S\|)
%\label{measure_fun3}
%\\
%[\mathbf g(\mathbf c_{k|k-1,p(k)})]_4 =& \frac{1}{\lambda} \left( \frac{(\mathbf v_{k,p(k)})^T (\mathbf p_{k,p(k)} - \mathbf p_B)} {\|\mathbf v_{k,p(k)}\| \cdot \|\mathbf p_{k,p(k)} - \mathbf p_B\| } \!+\!
%\frac{(\mathbf v_{k,p(k)})^T (\mathbf p_{k,p(k)} - \mathbf p_S)} {\|\mathbf v_{k,p(k)}\| \cdot \|\mathbf p_{k,p(k)} - \mathbf p_S\|  }  \right)
%\label{measure_fun4}
\end{align}
where $\mathbf e_X$, $\mathbf e_Y$, $\mathbf e_Z$ are the unit vectors towards the three axes of the global coordinate system, respectively.

Combining \eqref{AR_model} and \eqref{AR_model_meas}, the nonlinear AR model for the dynamic targets can be constructed.
We then resort to GM-PHD filter for the tracking of multiple targets.

%\subsection{Multiple Target Tracking Problem formulation}

\subsection{GM-PHD Filter}
Since the number of targets is variational, traditional target tracking methods cannot be directly applied to address this problem.
However, the concept of random finite set (RFS) matches the requirement.
Since the states of the targets belong to a finite set with a specific range of positions, it can be checked that the set of the states is a RFS \cite{RFS}.
Moreover, it is convenient to add or remove elements with the RFS, which means it can properly deal with the variation of the target amounts.
Thus, the aim of the multi-target tracking problem is to track the variation of the RFS.
Note that the order in which the states are listed has no significance in the RFS multi-target model formulation.

With the help of RFS, we resort to GM-PHD filter for tracking a variational number of targets.
There are three steps in the GM-PHD filter \cite{GM_PHD1}: prediction and updating, pruning and merging, and state extraction.
Prediction and updating step gives the intensities of Gaussian components in the next duration.
Pruning and merging step will remove the intensities with low amplitude and combine the intensities with close means.
Finally, the tracked target states will be acquired in the state extraction step.

\subsubsection{\bf Gaussian Intensity Prediction and Updating}
Assume that the posterior intensity $\boldsymbol v_{k-\!1}$ for the states of the targets at the $(k\!-\!1)$-th period is the mixture of $P(k-1)$ Gaussian components as follows:
\begin{align}\label{GM_intensity_fun}
\boldsymbol v_{k-1}(\mathbf c) = \sum_{j=1}^{P(k-1)} \omega_{k-1}^{(j)} \mathcal N (\mathbf c; \mathbf m_{k-1}^{(j)}, \mathbf P_{k-1}^{(j)}),
\end{align}
where $\omega_{k-1}^{(j)}$, $\mathbf m_{k-1}^{(j)}$, and $\mathbf P_{k-1}^{(j)}$ $(j = 1,2,\ldots, P(k-1))$ are respectively the weight, mean, and covariance matrix of the $j$-th Gaussian component in $\boldsymbol v_{k-1}$.
$P(k-1)$ is the number of the Gaussian components.

Then, we can give the predicted intensity at the $k$-th period:
\begin{align}\label{GM_intensity_fun_predict}
\boldsymbol v_{k|k-1} (\mathbf c) =
\boldsymbol v_{\gamma,k|k-1} (\mathbf c)
+ \boldsymbol v_{\kappa,k|k-1} (\mathbf c)
+ \boldsymbol v_{b, k} (\mathbf c),
\end{align}
where $\boldsymbol v_{\gamma,k|k-1} (\mathbf c)$, $\boldsymbol v_{\kappa,k|k-1} (\mathbf c) $, and $\boldsymbol v_{b, k} (\mathbf c)$ are respectively the surviving intensity, spawning intensity, and the birth intensity as follows:
\begin{align}
\boldsymbol v_{\gamma, k|k-\!1}(\!\mathbf c) \!\!=& p_{S,k} \!\!\sum_{j=1}^{P(k-1)} \!\!\! \omega_{k-1}^{(j)}
\mathcal N(\mathbf c; \mathbf m_{\gamma,k|k-1}^{(j)}, \mathbf P_{\gamma, k|k-1}^{(j)}),
\label{GM_intensity_fun_sur}
\\
\mathbf m_{\gamma,k|k-1}^{(j)} =& \mathbf A \mathbf m_{k-1}^{(j)},
\label{GM_intensity_mean_sur}
\\
\mathbf P_{\gamma, k|k-1}^{(j)} =& \mathbf Q_{j} + \mathbf A \mathbf P_{k-1}^{(j)}\mathbf A^T,
\label{GM_intensity_cov_sur}
\\
\boldsymbol v_{\kappa, k|{k\!-\!1}}(\mathbf c) \!\!=&\!\!\!\!\!\! \sum_{j=1}^{P(k\!-\!1)} \!\! \sum_{l=1}^{P_{\kappa}(k)}\!\!\! \omega_{k-\!1}^{(j)} \omega_{\kappa,k}^{(l)}
\mathcal N(\mathbf c; \!\mathbf m_{\kappa,k|k-\!1}^{(j,l)}, \!\mathbf P_{\kappa, k|k-\!1}^{(j,l)}\!),
\label{GM_intensity_fun_spawn}
\\
\mathbf m_{\kappa,k|k-1}^{(j,l)} =& \mathbf A \mathbf m_{k-1}^{(j)}+ \mathbf e_{\kappa,k-1}^{(l)},
\label{GM_intensity_mean_spawn}
\\
\mathbf P_{\kappa, k|k-1}^{(j,l)} =& \mathbf Q_{j} + \mathbf A \mathbf P_{k-1}^{(j)}\mathbf A^T,
\label{GM_intensity_cov_spawn}
\\
\boldsymbol v_{b,k} (\mathbf c) =& \sum_{i=1}^{P_{b}(k)} \omega _{b,k}^{(i)} \mathcal N(\mathbf c; \mathbf m_{b,k}^{(i)}, \mathbf P_{b,k}^{(i)}),
\label{GM_intensity_fun_bir}
\end{align}
where $\mathbf m_{\gamma,k|k-1}^{(j)}, \mathbf P_{\gamma,k|k-1}^{(j)}$ with $j = 1,2,\ldots, P(k-1)$ determine the shape of the predicted surviving intensity.
Similarly, $P_{\kappa}(k)$ and $\omega_{\kappa,k}^{(l)}, \mathbf m_{\kappa,k|k-1}^{(j,l)}, \mathbf P_{\kappa,k|k-1}^{(j,l)}$ where $l = 1,2,\ldots, P_{\kappa}(k)$, determine the shape of the predicted spawning intensity.
Besides, the number $P_{\kappa}(k)$ and weight $\omega_{\kappa,k}^{(l)}$ of spawning intensities in the $k$-th period are determined by the type of the target, where the target classification has been illustrated in Sec. III-A.
Moreover, $P_{b}(k)$ and $\omega_{b,k}^{(i)}, \mathbf m_{b,k}^{(i)}, \mathbf P_{b,k}^{(i)}$ are the given model parameters that determine the shape of predicted birth intensity, and $i = 1,2,\ldots, P_{b}(k)$.
Note that these parameters are determined by the environment.
In addition, $ p_{S,k}$ is the surviving probability of existed targets in the $k$-th period, and $\mathbf e_{\kappa,k-1}^{(l)}$ is a random deviation value.
With the above calculation, the predicted number of targets  in the $k$-th period is
$\widetilde{P}(k|k-1) = P(k-1) + P(k-1)P_{\kappa}(k) + P_b(k)$.

Therefore, the predicted posterior intensity at the $k$-th period is also Gaussian mixture, and \eqref{GM_intensity_fun_predict} can be rewritten as
\begin{align}
\boldsymbol v_{k|k-1} (\mathbf c) =& \sum_{j=1}^{\widetilde{P}(k|k-1)}
\omega_{k|k-1}^{(j)}
\mathcal N(\mathbf c; \mathbf m_{k|k-1}^{(j)}, \mathbf P_{k|k-1}^{(j)}),
\label{predicted_posterior_GM_intensity_fun}
\end{align}
where $\mathbf m_{k|k-1}^{(j)}$ and $\mathbf P_{k|k-1}^{(j)}$ are the united sets of
$\{\mathbf m_{\gamma,k|k-1}^{(j)}, \mathbf m_{\kappa,k|k-1}^{(j,l)}, \mathbf m_{b,k}^{(i)}\}$ and
$\{\mathbf P_{\gamma, k|k-1}^{(j)}, \mathbf P_{\kappa, k|k-1}^{(j,l)}, \mathbf P_{b,k}^{(i)}\}$, respectively.
Furthermore, the updated intensity at the $k$-th period can be represented as follows:
\begin{align}
\widetilde{\boldsymbol v}_{k} (\mathbf c) =& (1-\!p_{D,k}) \boldsymbol v_{k|k-1}(\mathbf c) \!+ p_{D,k} \!\!\!\!\sum_{\mathbf z_{k,l} \in \mathcal Z_k} \!\!\!\!\boldsymbol v_{D,k}(\mathbf c; \mathbf z_{k,l}),
\label{updated_and_posterior_GM_intensity_fun}
\end{align}
where $p_{D,k}$ is the detection probability, representing the probability that the observation set can match the target set.
Moreover, $\mathbf z_{k,l}$ is the $l$-th element in $\mathcal Z_k$.
Especially, the second item at the right hand side of \eqref{updated_and_posterior_GM_intensity_fun} denotes the intensity function with the update by using the observation set $\mathcal Z_k$,
which is represented as
\begin{align}
\boldsymbol{v}_{D, k}(\mathbf c ; \mathbf z_{k,l}) \!\!=& \!\!\!\!\! \sum_{j=1}^{\widetilde{P}(k|k-1)} \!\!\!\! \omega_k^{(j)}(\mathbf z_{k,l})
\mathcal N \!\left(\mathbf c; \mathbf m_{k |k}^{(j)} (\mathbf z_{k,l}), \mathbf P_{k |k}^{(j)}\right),
\label{predicted_GM_intensity_para_fun1}
\end{align}
for $l=1,2,\ldots, \text{size}(\mathcal Z_k)$, where
\begin{align}
\omega_k^{(j)}\!(\mathbf z_{k\!,l\!}) \!=&\! \frac{ \omega_{k | k-1}^{(j)}
\mathcal N(\mathbf z_{k,l}; \mathbf g(\mathbf m_{k|k-1}^{(j)}\!), \mathbf S_k^{(j)}) }
{C\!_k\!(\!\mathbf {z}_{k\!,l}\!)\!/\!p_{D\!,k}  \!\!+\! \!\!\!\!\!\!\!\sum\limits_{l=1}^{\widetilde{P}(k|k\!-\!1)} \!\!\!\!\!\!\! w_{k | k\!-\!1}^{(l)} \!\mathcal N \!(\!\mathbf z_{k\!,l}; \!\mathbf g(\!\mathbf m_{k|k\!-\!1}^{(l)}\!), \!\mathbf S\!_k^{(\!l)}\!) },
\label{predicted_GM_intensity_para_fun2}
\\
\mathbf m_{k|k}^{(j)}\!(\mathbf z_{k,l}\!) \!=& \mathbf m_{k | k-1}^{(j)} + \mathbf K_k^{(j)} \left(\mathbf z_{k,l}- \mathbf g(\mathbf m_{k | k-1}^{(j)})\right),
\label{predicted_GM_intensity_para_fun3}
\\
\mathbf P_{k|k}^{(j)} =& \left(\mathbf I - \mathbf K_k^{(j)} \mathbf G_k^{(j)}(\mathbf m_{k|k\!-1}^{(j)}\!) \right) \mathbf P_{k | k-1}^{(j)},
\label{predicted_GM_intensity_para_fun4}
\end{align}
determine the shape of the updated posterior intensity.
Moreover, $C_k(\mathbf {z}_{k,l}) $ in \eqref{predicted_GM_intensity_para_fun2} is the clutter intensity, which is usually assumed to be a Poisson distributed random finite set, i.e.,
$
C_k(\mathbf z_{k,l}) = \lambda_{\text {clutter }} V_s U(\mathbf z_{k,l})
$,
where $\lambda_{\text {clutter }}$ is the clutter rate, $V_s$ refers to the volume of the measurement region, and $U(\cdot)$ is a uniform density.
Besides, $\mathbf G_k^{(j)}(\mathbf m_{k|k\!-1}^{(j)}\!)$ is the derivative matrix of $\mathbf g(\mathbf m_{k|k\!-1}^{(j)}\!)$ with respect to $\mathbf m_{k|k-1}^{(j)}$.
In addition,
\begin{align}
\mathbf K_k^{(j)} =& \mathbf P_{k |k-1}^{(j)} (\mathbf G_k^{(j)}(\mathbf m_{k|k\!-1}^{(j)}\!))^T \left(\mathbf S_k^{(j)}\right)^{-1},
\label{predicted_GM_intensity_para_fun5}
\\
\mathbf S_k^{(j)} =& \!\mathbf G_k^{(j)}\!(\mathbf m_{k|k\!-1}^{(j)}\!) \mathbf P_{k | k\!-1}^{(j)} (\!\mathbf G_k^{(j)}\!(\mathbf m_{k|k\!-1}^{(j)}\!))^T \!\!\!+\! \mathbf R_{k,j}.
\label{predicted_GM_intensity_para_fun6}
\end{align}

Thus, \eqref{updated_and_posterior_GM_intensity_fun} has the similar form with \eqref{GM_intensity_fun}, and contains $\widetilde{P}(k) = \widetilde{P}(k|k-1) + \text{size}(\mathcal Z_k) \widetilde{P}(k|k-1)$ components, i.e.,
\begin{align}\label{updated_GM_intensity_fun}
\widetilde{\boldsymbol v}_{k}(\mathbf c) = \sum_{j=1}^{\widetilde{P}(k)} \widetilde{\omega}_{k}^{(j)} \mathcal N (\mathbf c; \widetilde{\mathbf m}_{k}^{(j)}, \widetilde{\mathbf P}_{k}^{(j)}).
\end{align}
Especially, for $j=1,2,\ldots, \widetilde{P}(k|k-1)$, we have $\widetilde{\omega}_k^{(j)} = (1-p_{D,k}) \omega_{k|k-1}^{(j)}$,
$\widetilde{\mathbf m}_k^{(j)} = \mathbf m_{k|k-1}^{(j)}$,
and $\widetilde{\mathbf P}_k^{(j)} = \mathbf P_{k|k-1}^{(j)}$.
Moreover, for $l = 1,2, \ldots, \text{size}(\mathcal Z_k)$ and $j=1,2,\ldots, \widetilde{P}(k|k-1)$, we have
$\widetilde{\omega}_k^{(l\widetilde{P}(k|k-1)+j)} = p_{D,k} \omega_k^{(j)}(\mathbf z_{k,l})$,
$\widetilde{\mathbf m}_k^{(l\widetilde{P}(k|k-1)+j)} \!=\! \mathbf m_{k|k}^{(j)}(\mathbf z_{k,l})$,
and $\widetilde{\mathbf P}_k^{(l\widetilde{P}(k|k-1)+j)} \!=\! \mathbf P_{k|k}^{(j)}$.
The processes of prediction and updating are summarized in {\bf Algorithm \ref{alg:prediction}}.

\begin{algorithm}
	\caption{Gaussian intensity prediction and updating}
	\label{alg:prediction}
	\renewcommand{\arraystretch}{0.9}
    %\setstretch{1.35}
	\begin{algorithmic}[1]
        \STATE {\bf Input:} $\{\omega_{k-1}^{(i)}, \mathbf m_{k-1}^{(i)}, \mathbf P_{k-1}^{(i)}\}_{i=1}^{P(k-1)}$, and $\mathcal Z_k$.
        \\
        {\bf(prediction step:)}
        \STATE For the survived targets, the predicted parameters in the intensity can be calculated as \eqref{GM_intensity_fun_sur}-\eqref{GM_intensity_cov_sur}.
        \STATE For the spawned targets, the predicted parameters in the intensity can be calculated as \eqref{GM_intensity_fun_spawn}-\eqref{GM_intensity_cov_spawn}.
        \STATE For the birth targets, the predicted parameters in the intensity can be generated as \eqref{GM_intensity_fun_bir}.
        \STATE $\widetilde{P}(k|k-1) = P(k-1) + P(k-1)P_{\kappa}(k) + P_b(k)$.
        \\
        {\bf(PHD update components construction)}
        \FOR {$j = 1, 2, \cdots, \widetilde{P}(k|k-1)$}
        \STATE $\{\mathbf S_k^{(j)}, \mathbf K_k^{(j)}, \mathbf P_{k|k}^{(j)}\} \leftarrow$ calculate \eqref{predicted_GM_intensity_para_fun4}-\eqref{predicted_GM_intensity_para_fun6}.
        \FOR {$l = 1, 2, \cdots, \text{size}(\mathcal Z_k)$}
        \STATE $\{\omega_k^{(j)}(\mathbf z_{k,l}), \mathbf m_{k|k}^{(j)}(\mathbf z_{k,l})\} \leftarrow$ calculate \eqref{predicted_GM_intensity_para_fun2}-\eqref{predicted_GM_intensity_para_fun3}.
        \ENDFOR
        \ENDFOR
        \\
        {\bf(updating step:)}
        \FOR {$j = 1, 2, \cdots, \widetilde{P}(k|k-1)$}
        \STATE $\widetilde{\omega}_k^{(j)} \!\!=\! (1\!-\!p_{D,k}) \omega_{k|k-1}^{(j)}$,
        $\widetilde{\mathbf m}_k^{(j)} \!\!=\!\! \mathbf m_{k|k-1}^{(j)}$,
        $\widetilde{\mathbf P}_k^{(j)} \!\!=\!\! \mathbf P_{k|k-1}^{(j)}$.
        \ENDFOR
        \FOR {$l = 1, 2, \ldots, \text{size} (\mathcal Z_k)$}
        \FOR {$j = 1, 2, \cdots, \widetilde{P}(k|k-1)$}
        \STATE $\widetilde{\omega}_k^{(l\widetilde{P}(k|k-1)+j)} = \omega_k^{(j)}(\mathbf z_{k,l})$,
        $\widetilde{\mathbf m}_k^{(l\widetilde{P}(k|k-1)+j)} = \mathbf m_{k|k}^{(j)}(\mathbf z_{k,l})$,
        $\widetilde{\mathbf P}_k^{(l\widetilde{P}(k|k-1)+j)} = \mathbf P_{k|k}^{(j)}$.
        \ENDFOR
        \ENDFOR
        \STATE $\widetilde{P}(k) = \widetilde{P}(k|k-1) + \text{size}(\mathcal Z_k) \widetilde{P}(k|k-1)$.
        \RETURN {$\widetilde{P}(k)$, $\{\widetilde\omega_k^{(j)}, \widetilde{\mathbf m}_k^{(j)}, \widetilde {\mathbf P}_k^{(j)}\}_{j=1}^{\widetilde{P}(k)}$. } 	
	\end{algorithmic}
\end{algorithm}

\subsubsection{\bf Gaussian Component Pruning and Merging}

With the processing of the algorithm, the number of Gaussian components increases exponentially, leading to a high computational complexity \cite{GM_PHD_pruning}.
Thus, the number of Gaussian components propagated to the next period should be reduced by pruning operation.
The pruning can be implemented by truncating components in \eqref{updated_GM_intensity_fun} that have weak weights $\widetilde\omega_k^{(j)}$.
This can be achieved by leaving out the components with lower weights than a preset threshold $\zeta$, or reserve a certain number of components with stronger weights that occupy a sufficient portion of the total weights of all the components.
Define $\mathcal I$ as the set for the indexes of the remained components after pruning for further use, which can be derived as $\mathcal J = \{j = 1, \ldots, \widetilde{P}(k) | \widetilde\omega_k^{(j)} > \zeta \} $.

Moreover, the states of some Gaussian components may be so close, thus they could be approximated and merged into a single Gaussian component.
Define the merging threshold $\xi$, and set $i$ as the index of the Gaussian component in $\mathcal J$ with the highest weight.
Then the set $\mathcal L$ for the indexes of the components to be merged into a single component can be derived as
\begin{align}
\mathcal L\!\! = \!\!\left\{j \!\in\! \mathcal J \mid \!(\widetilde{\mathbf m}_k^{(j)} \!\!-\! \widetilde{\mathbf m}_k^{(i)})^T (\widetilde{\mathbf P}_k^{(j)})^{-\!1} (\widetilde{\mathbf m}_k^{(j)} \!\!-\! \widetilde{\mathbf m}_k^{(i)}) \!\leq\! \xi\right\}.
\end{align}
Correspondingly, the parameters of the $\ell$-th merged component can be calculated as
\begin{align}
\overline \omega_k^{(\ell)} \!\!=& \sum\limits_{j\in \mathcal L} \widetilde\omega_k^{(j)},
\label{merge_compo_para1}
\\
\overline{\mathbf m}_k^{(\ell)} \!\!=& \frac{1}{\overline \omega_k^{(\ell)}} \sum\limits_{j\in \mathcal L} \widetilde\omega_k^{(j)} \widetilde{\mathbf m}_k^{(j)},
\label{merge_compo_para2}
\\
\overline{\mathbf P}_k^{(\ell)} \!\!=&\! \frac{1}{\overline \omega_k^{(\ell)}} \!\!\sum\limits_{j\in \mathcal L} \!\widetilde\omega_k^{(j)} (\widetilde{\mathbf P}_k^{(j)} \!\!+\! (\overline{\mathbf m}_k^{(\ell)} \!\!-\! \widetilde{\mathbf m}_k^{(j)}) (\overline{\mathbf m}_k^{(\ell)} \!\!\!-\! \widetilde{\mathbf m}_k^{(j)})^T ).
\label{merge_compo_para3}
\end{align}
Then, the elements of $\mathcal L$ are removed from $\mathcal J$.
By iteratively implementing the above operations, we can empty $\mathcal J$ and obtain a number of merged Gaussian components.
Finally, $P_{\text{max}}$ Gaussian components with largest weights are picked up from the merged ones.
Note that the merging is just to mathematically deal with the Gaussian components rather than actual targets.
Thus, the updated posterior intensity after pruning and merging can be represented as
\begin{align}\label{updated_GM_intensity_fun_pruned_merged}
\overline{\boldsymbol v}_{k}(\mathbf c) = \sum_{j=1}^{P_{\text{max}}} \overline{\omega}_{k}^{(j)} \mathcal N (\mathbf c; \overline{\mathbf m}_{k}^{(j)}, \overline{\mathbf P}_{k}^{(j)}).
\end{align}
The processes of pruning and merging are summarized in {\bf Algorithm \ref{alg:Pruning}}.

\begin{algorithm}
	\caption{Pruning and merging for the GM-PHD filter}
	\label{alg:Pruning}
	\renewcommand{\arraystretch}{0.9}
    %\setstretch{1.35}
	\begin{algorithmic}[1]
        \STATE {\bf Input:} $\{\widetilde{\omega}_k^{(j)}, \widetilde{\mathbf m}_k^{(j)}, \widetilde{\mathbf P}_k^{(j)}\!\}_{j=1}^{\widetilde{P}(k)}$, the truncation threshold $\zeta$, and the merging threshold $\xi$.
        \STATE {\bf Initialize:} $\ell = 0$, and $\mathcal J = \{j = 1, \ldots, \widetilde{P}(k) | \widetilde{\omega}_k^{(j)} > \zeta \} $.
        \WHILE {$\mathcal J \neq \emptyset$}
        \STATE $\ell = \ell +1$.
        \STATE $i = \arg \max\limits_{i\in \mathcal I} \widetilde\omega_k^{(i)}$.
        \STATE $\mathcal L\! = \!\!\left\{j \!\in\! \mathcal J \mid \!(\widetilde{\mathbf m}_k^{(j)} \!\!-\! \widetilde{\mathbf m}_k^{(i)})^T (\widetilde{\mathbf P}_k^{(j)})^{-\!1} (\widetilde{\mathbf m}_k^{(j)} \!-\! \widetilde{\mathbf m}_k^{(i)}) \!\leq\! \xi\right\}$.
        \STATE  $\{\overline \omega_k^{(\ell)}, \overline{\mathbf m}_k^{(\ell)}, \overline{\mathbf P}_k^{(\ell)}\} \leftarrow$ calculate \eqref{merge_compo_para1}-\eqref{merge_compo_para3}.
%        \STATE $\widetilde \omega_k^{(l)} = \sum\limits_{i\in \mathcal L} \omega_k^{(i)}$.
%        \STATE $\widetilde{\mathbf m}_k^{(l)} = \frac{1}{\widetilde \omega_k^{(l)}} \sum\limits_{i\in \mathcal L} \omega_k^{(i)} \mathbf m_k^{(i)}$.
%        \STATE $\widetilde{\mathbf P}_k^{(l)} \!\!=\! \frac{1}{\widetilde \omega_k^{(l)}} \!\sum\limits_{i\in \mathcal L} \!\omega_k^{(i)} (\mathbf P_k^{(i)} \!+\! (\widetilde{\mathbf m}_k^{(l)} \!-\! \mathbf m_k^{(j)}) (\widetilde{\mathbf m}_k^{(l)} \!-\! \mathbf m_k^{(j)})^T )$.
        \STATE $\mathcal J = \mathcal J \setminus \mathcal L$.
        \ENDWHILE
        \IF {$\ell > P_{\text{max}}$}
        \STATE {Replace $\{\overline\omega_k^{(j)}, \overline{\mathbf m}_k^{(j)}, \overline {\mathbf P}_k^{(j)}\}_{i=1}^{\ell}$ by picking up $P_{\text{max}}$ Gaussian components with largest weights}.
        \ELSE
        \STATE {$P_{\text{max}} = \ell$}
        \ENDIF
        \RETURN {$\{\overline\omega_k^{(j)}, \overline{\mathbf m}_k^{(j)}, \overline {\mathbf P}_k^{(j)}\}_{j=1}^{P_{\text{max}}}$. } 	
	\end{algorithmic}
\end{algorithm}

\subsubsection{\bf Multi-target State Extraction}

After obtaining the posterior intensity $\overline{\boldsymbol v}_{k}(\mathbf c)$, the states of multi-targets can be directly extracted by searching the peaks of $\overline{\boldsymbol v}_{k}(\mathbf c)$.
However, the peaks in $\overline{\boldsymbol v}_{k}(\mathbf c)$ are related to not only the weights but also the covariance of all the Gaussian components.
Thus, choosing a number of highest peaks directly in $\overline{\boldsymbol v}_{k}(\mathbf c)$ may result in the selection of weaker Gaussian components, and cause the unexpected tracking results.
Alternatively, we can choose the components that have weights larger than a threshold $\rho$ as the final result of the tracked multi-target states.
Then, the set of the tracked target states $\mathcal C_k$ can be obtained, and the estimated number of the targets can be calculated by $\widehat{P}(k) = \text{size}(\mathcal C_k)$.
The process of multi-target state extraction is summarized in {\bf Algorithm \ref{alg:state_extraction}}.

\begin{algorithm}
	\caption{Multi-target state extraction}
	\label{alg:state_extraction}
	\renewcommand{\arraystretch}{0.9}
    %\setstretch{1.35}
	\begin{algorithmic}[1]
        \STATE {\bf Input:} $\{\overline\omega_k^{(j)}, \overline{\mathbf m}_k^{(j)}, \overline {\mathbf P}_k^{(j)}\}_{j=1}^{P_{\text{max}}}$.
        \STATE {\bf Initialize:} $\mathcal C_k = \emptyset$.
        \FOR {$j = 1, \cdots, {P_{\text{max}}}$}
        \IF {$\overline\omega_k^{(j)} > \rho$}
        \STATE $\mathcal C_k = \mathcal C_k \cup \{\overline{\mathbf m}_k^{(j)}\}$.
        \ENDIF
        \ENDFOR
        \STATE $\widehat{P}(k) = \text{size}(\mathcal C_k)$.
        \RETURN {$\widehat{P}(k)$, $\mathcal C_k$. } 	
	\end{algorithmic}
\end{algorithm}

By successively implementing {\bf Algorithm \ref{alg:prediction}, \ref{alg:Pruning} and \ref{alg:state_extraction}}, we can obtain the tracked state of multiple targets.

%Combining all the processes above, the overall procedure of the GM-PHD filter is summarized in {\bf Algorithm \ref{alg:GM_PHD_overall_procedure}}.
%
%\begin{algorithm}
%	\caption{The overall procedure of GM-PHD filter}
%	\label{alg:GM_PHD_overall_procedure}
%	\renewcommand{\arraystretch}{0.9}
%    %\setstretch{1.35}
%	\begin{algorithmic}[1]
%        \STATE {\bf Input:} $\{\omega_{k-1}^{(i)}, \mathbf m_{k-1}^{(i)}, \mathbf P_{k-1}^{(i)}\}_{i=1}^{P(k-1)}$, and the measurement set $\mathcal Z_k$.
%        \\
%        {\bf(prediction and updating step:)}
%        \STATE $\widetilde{P}(k)$, $\{\widetilde\omega_k^{(j)}, \widetilde{\mathbf m}_k^{(j)}, \widetilde {\mathbf P}_k^{(j)}\}_{j=1}^{\widetilde{P}(k)}$ $\leftarrow$ \textbf{Algorithm \ref{alg:prediction}}.
%        \\
%        {\bf(pruning and merging step:)}
%        \STATE $\{\overline\omega_k^{(j)}, \overline{\mathbf m}_k^{(j)}, \overline{\mathbf P}_k^{(j)}\}_{j=1}^{P_{\text{max}}}$ $\leftarrow$ \textbf{Algorithm \ref{alg:Pruning}}.
%        \\
%        {\bf(state extraction step:)}
%        \STATE $\widehat{P}(k), \mathcal C_k$ $\leftarrow$ \textbf{Algorithm \ref{alg:state_extraction}}.
%        \RETURN {$\widehat P(k)$, $\mathcal C_k$. } 	
%	\end{algorithmic}
%\end{algorithm}

\section{Communication Performance Optimization}

\subsection{BS Beamforming and STAR-RIS Phase Shift Prediction via Tracked Scattering Paths}
After the extraction of multiple target states, we aim to improve the transmission quality by delicately adjusting the beamforming of the BS and phase shift of the STAR-RIS in the subsequent data transmission phases.
Since the positions of all the possible targets $\{ \widehat{\mathbf p}_{k,p}\}_{p = 1}^{\widehat{P}(k)}$ have been predicted in the sensing period, we can directly derive the predicted channel parameters, and further develop the beamforming of the BS and the phase shift of the STAR-RIS for performance improvement.

Generally, the design of the beam depends on the geometrical parameters between the nodes in communication system,
thus we focus on the reconstruction of the AODs at the BS and the AOAs at the STAR-RIS.
Take the $p$-th predicted target with the predicted position $\widehat{\mathbf p}_{k,p}$ as an example.
Since the position of BS and the STAR-RIS is known in a priori, the predicted direction of the $p$-th predicted target with respect to the BS and the STAR-RIS can be derived as $\widehat{\mathbf p}_{k,p} - \mathbf p_B$ and $\widehat{\mathbf p}_{k,p} - \mathbf p_S$, respectively.
By replacing ${\mathbf p}_{k,p}$ with $\widehat{\mathbf p}_{k,p}$ in \eqref{measure_fun1}-\eqref{measure_fun2}, the predicted AODs at BS can be calculated as
\begin{align}\label{predic_angles_BS}
\widehat{\phi}_{p}^B (k) \!=&
%\arccos \frac{\Big((\widehat{\mathbf p}_{k,p(k)} - \mathbf p_B) - \frac{(\widehat{\mathbf p}_{k,p(k)} - \mathbf p_B)^T \mathbf e_{Y}}{\|\mathbf e_{Y}\|} \mathbf e_{Y}\Big)^T \mathbf e_{X}}
%{\big\|(\widehat{\mathbf p}_{k,p(k)} - \mathbf p_B) - \frac{(\widehat{\mathbf p}_{k,p(k)} - \mathbf p_B)^T \mathbf e_{Y}}{\|\mathbf e_{Y}\|} \mathbf e_{Y}\big\| \cdot \left\|\mathbf e_{X}\right\|},
[\mathbf g(\widehat{\mathbf c}_{k,p})]_1,
\quad\quad\quad
\widehat{\psi}_{p}^B (k) \!=
%\frac{\pi}{2} - \arccos\frac{(\widehat{\mathbf p}_{k,p(k)} - \mathbf p_B)^T \mathbf e_{X}}{\|\widehat{\mathbf p}_{k,p(k)} - \mathbf p_B\| \cdot \|\mathbf e_{X}\|},
[\mathbf g(\widehat{\mathbf c}_{k,p})]_2.
\end{align}
for $p = 1, 2, \ldots, \widehat{P}(k)$.
Similarly, the predicted AOAs at the STAR-RIS can be further written as
\begin{align}\label{predic_angles_RIS}
\widehat{\phi}_{p}^S (k) \!=&\!\arccos \!\frac{\big((\widehat{\mathbf p}_{k,p} \!-\!\! \mathbf p_S) \!-\! \frac{(\widehat{\mathbf p}_{k,p} - \mathbf p_S)^T\mathbf e_Y}{\|\mathbf e_Y\|^2}\mathbf e_Y \!\big)^T \!\!\mathbf e_X}
{\big\|\!(\widehat{\mathbf p}_{k,p} \!-\! \mathbf p_S) \!-\! \frac{(\widehat{\mathbf p}_{k,p}\!\! -\! \mathbf p_S)^T\mathbf e_Y}{\|\mathbf e_Y\|^2} \!\mathbf e_Y \!\big\| \!\cdot\! \|\! \mathbf e_X \! \|},
\\
\widehat{\psi}_{p}^S (k) \!=&
\frac{\pi}{2} - \arccos \frac{(\widehat{\mathbf p}_{k,p} - \mathbf p_S)^T \mathbf e_Y}{\|\widehat{\mathbf p}_{k,p} - \mathbf p_S\| \cdot \|\mathbf e_Y\|},
\end{align}
for $p = 1, 2, \ldots, \widehat{P}(k)$, respectively.
With the predicted angular information, we then propose a strategy for the design of the BS beamforming and STAR-RIS phase shifts.
Our idea is to firstly design the sub-beam for each scattering path, and then combine these sub-beams to form a final beam solution at the BS and the STAR-RIS, respectively.
Specifically, the beamforming for the $p$-th target at the BS and the refraction phase shift vector at the STAR-RIS can be respectively optimized by
\begin{align}\label{predic_beams_problem}
\overline{\mathbf f}\!_{B,p}(k) =&\!
\arg\max\limits_{\mathbf f_{B,p}}
\left|
\mathbf a_B^T(\widehat{\phi}_{p}^B (k), \widehat{\psi}_{p}^B (k))
 \mathbf f_{B,p}
\right|,
\\
\overline{\boldsymbol \omega}_{p} (\!k) \!\!=&
\arg\max\limits_{\boldsymbol \omega_{p}}
\left|
\mathbf a_R^T(\widehat\phi^S, \widehat\psi^S)\right.
\notag\\
&\left.\times\text{diag}({\boldsymbol \omega}_{p})
\mathbf{a}_O(\widehat{\phi}_{p}^S (k), \widehat{\psi}_{p}^S (k))
\right|,
\end{align}
where $\{\widehat\phi^S, \widehat\psi^S \}$ is the estimation of $\{\phi^S, \psi^S \}$, respectively.
Then the normalized solutions of $\overline{\mathbf f}_{B,p}(k)$ and $\overline{\boldsymbol \omega}_{p} (k) $ are respectively given by
\begin{align}\label{predic_beams}
\overline{\mathbf f}_{B,p}(k) =&
\mathbf a_B^*(\widehat{\phi}_{p}^B (k), \widehat{\psi}_{p}^B (k)),
\\
[\overline{\boldsymbol \omega}_{p}(k)]_{n_s} =&
\frac{[\mathbf{a}_{R}^* (\widehat{\phi}^S, \widehat{\phi}^S)]_{n_s}
[\mathbf{a}_{O}^* (\widehat{\phi}_{p}^S (k), \widehat{\psi}_{p}^S (k))]_{n_s} }
{\left|[\mathbf{a}_{R}^* (\widehat{\theta}^S, \widehat{\phi}^S)]_{n_s}
[\mathbf{a}_{O}^* (\widehat{\phi}_{p}^S (k), \widehat{\psi}_{p}^S (k))]_{n_s} \right|}.
\end{align}
where $n_s$ is the index of the element on the STAR-RIS.

Besides, the channel gain $\widehat{h}_{k,p}$ of the $p$-th predicted target is assumed to be equal to that of the target in the last sensing period.
Thus, by summing up and normalizing the designed phase shifts for all the predicted targets, the final predicted BS beamforming and the STAR-RIS refraction phase shift can be obtained as
\begin{align}\label{predic_beams_overall}
\overline{\mathbf f}_{B}\!(k) \!=\!&
\frac{\sum\limits_{p=1}^{\widehat{P}(k)}
\widehat{h}_{k,p}^* \overline{\mathbf f}_{B,p}(k)}
{\left\|\sum\limits_{p=1}^{\widehat{P}(k)}
\widehat{h}_{k,p}^* \overline{\mathbf f}_{B,p}(k)\right\|
}
,
\overline{\boldsymbol \omega}(k) \!=\!
\frac{\sum\limits_{p=1}^{\widehat{P}(k)} \widehat{h}_{k,p}^* \overline{\boldsymbol \omega}_{p}(k)}
{\left\|\sum\limits_{p=1}^{\widehat{P}(k)} \widehat{h}_{k,p}^* \overline{\boldsymbol \omega}_{p}(k)\right\|}.
\end{align}
Furthermore, the designed analog precoder of BS can be represented as
\begin{align}\label{predic_beams_overall_BSanalog}
\overline{\mathbf F}_{B}(k) =&
\frac{1}{\sqrt{N_B^{RF}}}
\overline{\mathbf f}_{B}(k) \mathbf 1_{N_B^{RF}}^T.
\end{align}

\begin{remark}
Since we focus on the tracking of the dynamic scatters in this paper, only the sub-beams for the tracked dynamic scatterers are illustrated in this subsection.
Note that there are also static scatterers in the environment.
Thus, the static sub-beams for those static scattering paths should also be incorporated in the overall predicted BS beamforming and STAR-RIS phase shift.
Therefore, \eqref{predic_beams_overall} will be further rebuilt with the consideration of the static scatters.
\end{remark}

\subsection{Target Mismatch Detection and Path Collision Prediction Mechanism Design}

As the time goes by, the tracked states of the targets may not match the actual states due to the accumulation of the tracking error resulting from inaccurate channel measurements.
If the beamforming of the BS and the phase shifts of the STAR-RIS are still developed by the mismatched states of the targets, the performance of data transmission will be deteriorated.
Under such circumstance, full-space beam scanning should restart.
Although we can periodically implement the full-space beam scanning, it will results in large training cost.
Thus, it is necessary and efficient to develop a mismatch detection mechanism to properly trigger the full-space beam scanning process with a maintained transmission performance.
On the other hand, with the moving of the target, it may obstruct the existing scattering paths passing through static scatterer between the BS and the STAR-RIS.
This will cause the collision between static scattering paths and the dynamic ones, and also affect the phase shifts design and prediction.
Thus, the collision of different scattering paths should be predicted.

\subsubsection{\bf Target Mismatch Detection}
\begin{figure}[htbp]
 \centering
 \includegraphics[width=85mm]{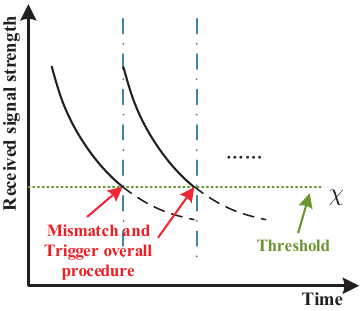}
 \caption{Illustration of the target mismatch detection mechanism.}
 \label{mismatch_detection}
\end{figure}

Here, we use the received signal strength at the indoor user as the criteria for target mismatch detection.
With respect to \eqref{UT_receive_nB} where $\mathbf F_B(k) \!\triangleq\! \overline{\mathbf F}_B (k)$ and $\boldsymbol \omega(k) \triangleq \overline{\boldsymbol \omega}(k)$,
it can be checked that the received signal strength $\eta_k = \|\overline{\mathbf y}_k^{U}\|^2$ will achieve its maximal value $\eta_{\max}$ when the phase shifts are optimized.
If the targets are not accurately tracked, the received signal strength $\|\overline{\mathbf y}_k^{U}\|^2$ will decrease.
Thus, we give a threshold $\chi$ of the received signal strength.
As illustrated in \figurename{ \ref{mismatch_detection}},
when $\eta_k < \chi$, the probability of mismatch becomes higher, and the full-space beam scanning should be triggered.

\subsubsection{\bf Path Collision Prediction}
For the path collision prediction, we resort to the location and derived angular information of the tracked targets as well as those of existing static scatterers.
If the predicted angular information of the targets, i.e.,
$\{\widehat{\phi}_{p}^B (k), \widehat{\psi}_{p}^B (k)\}_{p=1}^{\widehat{P}(k)}$ with respect to the BS or $\{\widehat{\phi}_{p}^S (k), \widehat{\psi}_{p}^S (k)\}_{p=1}^{\widehat{P}(k)}$ with respect to the STAR-RIS at the $k$-th ISAC period, is close to that of existing static scatterers, it can be determined that the path collision will occur at the $k$-th ISAC period.
Note that relative position between the target and its collided static scatterer can also be determined.
On the one hand, if the tracked target blocks the collided static scatterer, the sub-beam with respect to the collided static scatterer can be removed.
On the other hand, if the tracked target is blocked by the collided static scatterer, it will be unnecessary to design the sub-beam with respect to the target.
%On the one hand, if the tracked target is at the BS side of the collided scatterer, the sub-beam with respect to the collided scatterer can be removed.
%On the other hand, if the tracked target is at the STAR-RIS side of the collided scatterer, it will be unnecessary to design the sub-beam with respect to the target.
With the path collision prediction, the BS can focus the transmitted power on the useful scattering paths, which can improve the received signal power at the UT.
\begin{remark}
In practical communication system, the received signal strength of the indoor user in a specific time period can be utilized to determine whether the outdoor targets are mismatched or not.
For example, when the signal strength received by the indoor user gradually decreases and becomes lower than a pre-defined threshold, then it can be determined that the target mismatch exists.
For the prediction of target mismatch,
the acquired sensing information should be further utilized to analyze the tendency of the received signal strength of the indoor user.
If its received signal strength is a little higher than the threshold, and have the tendency of decreasing, then it can be predicted that target mismatching may occur in subsequent frames.
%For the prediction of target mismatch, we can also reuse the acquired sensing information to predict the path collision between the targets through the analysis of different target positions, and then predict the optimal beam to avoid target mismatch.
\end{remark}

\begin{figure}[htbp]
 \centering
 \includegraphics[width=90mm]{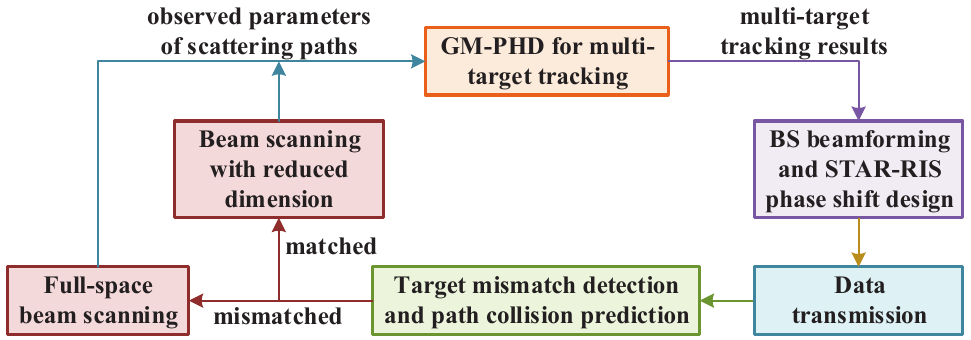}
 \caption{Overall procedure of the dynamic scatterer environment tracking enabled communication enhancement.}
 \label{flowing_fig_overall_tracking}
\end{figure}

The multi-target tracking scheme in the last section and the communication performance optimization in this section constitute the overall procedure of the proposed STAR-RIS assisted dynamic scatterers tracking scheme for ISAC, which is summarized in \figurename{ \ref{flowing_fig_overall_tracking}}.

\section{Simulation Results}

In this section, we evaluate the performance of our proposed tracking scheme for ISAC through numerical simulation.
Without loss of generality, we consider an interested area of $50$m $\times$ $50$m $\times$ $12$m.
The height of BS, STAR-RIS, and the dynamic scatterers are set as $12$m, $6$m, and $0$m, respectively.
The BS and the STAR-RIS are both equipped with $16\times 16$ UPA.
Besides, the number of RF chains $N_B^{RF}$ and $N_S^{RF}$ at the BS and the STAR-RIS are respectively set as $32$ and $5$.
The carrier frequency is $30$ GHz, and the antenna spacing $d$ is set as half of the wavelength.
The system sampling rate is $T_s = \frac{1}{100 \text{MHz}}$.
The maximal velocity of dynamic scatterers is set as $|v_{max}| = 3$ m/s, and the ISAC period is $0.5$ s.
The initial number of dynamic scatterers is $P(0) = 3$, and the scatterers are randomly placed in the sensing area with random velocities.
Besides, the number of dynamic scatterers entering the sensing area within the overall sensing process and those of dynamic scatterers spawning by existing scatterers are both set randomly.
Moreover, the exponentially decaying power delay profile $\sigma_{h_{p}}^2 = \sigma_c^2 e^{-\frac{\tau_p}{\tau_{max}}}$ is applied to $h_p^{O}$, where $\sigma_c^2$ is chosen to normalize the average cascaded channel power gains.
Here, we use root-mean-square-error (RMSE) to measure the sensing performance, which is defined as
\begin{align}
\text{RMSE}_{\mathbf x} \!=\! \sqrt{\mathbb E \{\|\widehat{\mathbf x} - \mathbf x\|^2\}},
\mathbf x = \mathbf c_{k,p}, \{\phi_p^B(k), \psi_p^B(k)\}.
\end{align}

\subsection{Numerical Results with Target Number Unchanged}

\begin{figure}[htbp]
 \centering
 \includegraphics[width=90mm]{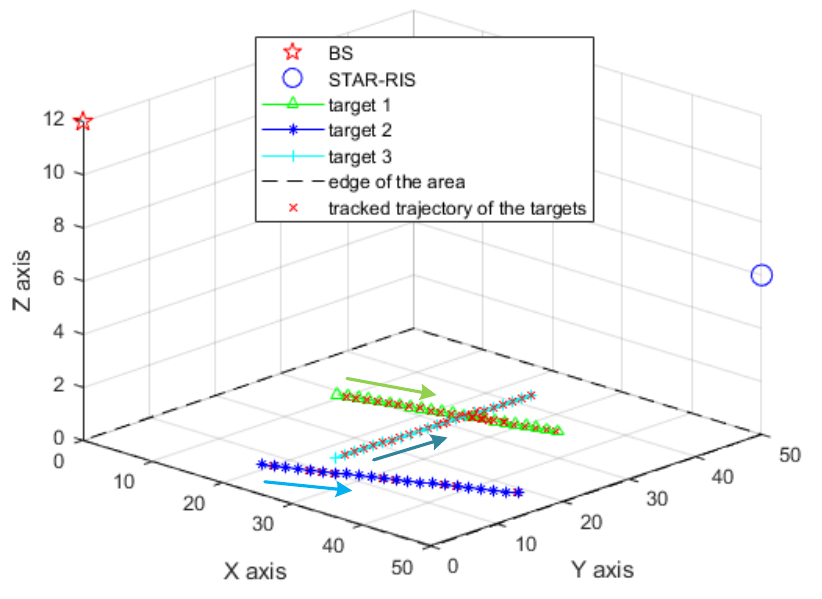}
 \caption{Illustration of multi-target tracking with fixed number of targets.}
 \label{scenario_illustration_tarInvariant}
\end{figure}

Firstly, we verify the proposed multi-target tracking scheme in the scenario that the number of the targets is fixed within the considered area and time duration.
The realization of multi-target tracking scenario and the tracking result are exhibited in \figurename{ \ref{scenario_illustration_tarInvariant}}, where the angle RMSE is set as $0.001^\circ$.
In \figurename{ \ref{scenario_illustration_tarInvariant}}, the black dashed curve is the observation area, the red star denotes the BS, and the bule circle denotes the STAR-RIS.
Besides, the green triangled curve, blue asterisk curve, and cyan cross curve respectively represent the actual trajectories of the three independent targets.
In addition, the red cross represents our tracking and predicting trajectories results.
It is obvious that the proposed GM-PHD for multi-target tracking scheme is effective and accurate.

\begin{figure}[htbp]
 \centering
 \includegraphics[width=95mm]{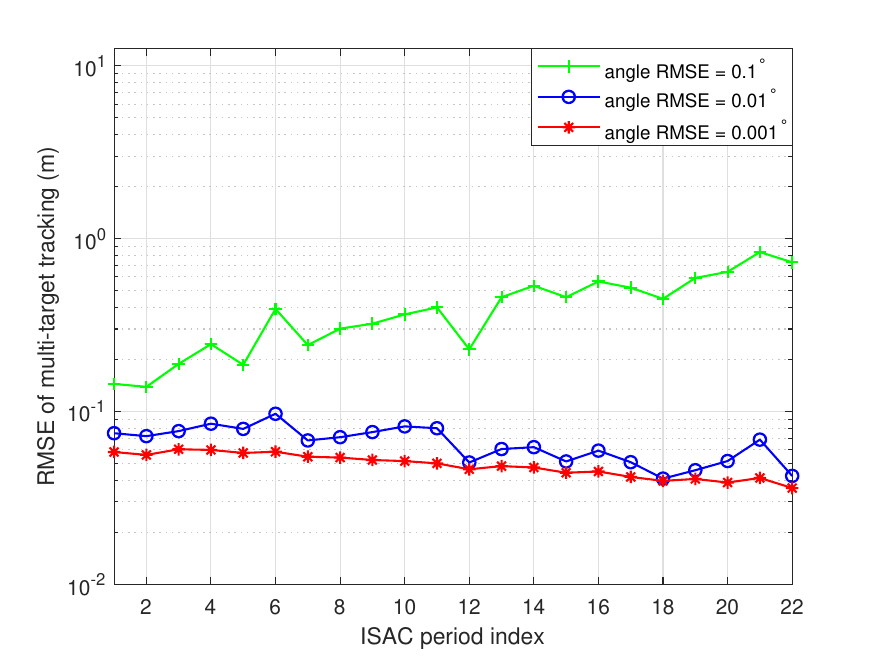}
 \caption{Multi-target tracking RMSEs versus ISAC period indexes under different level of angle errors ($v=3$m/s, 3 targets).}
 \label{locRMSE_TargetInvariation_vs_Timestep_differentAngleError}
\end{figure}

\figurename{ \ref{locRMSE_TargetInvariation_vs_Timestep_differentAngleError}} further illustrates the multi-target tracking RMSEs versus the ISAC period indexes under different levels of angle estimation errors.
It can be observed that the tracking results become more accurate with more precise angle estimation accuracy.
Besides, the tracking RMSE with lower angle RMSE is more stable, and that with higher angle RMSE has a larger fluctuation.
Moreover, the gap between the blue curve and the red one is significantly smaller than that between the blue curve and the green one.
This is due to the fact that the tracking performance turns to be dominated by the measurement error rather than algorithm error with the decrease of angle RMSE.

\begin{figure}[htbp]
 \centering
 \includegraphics[width=95mm]{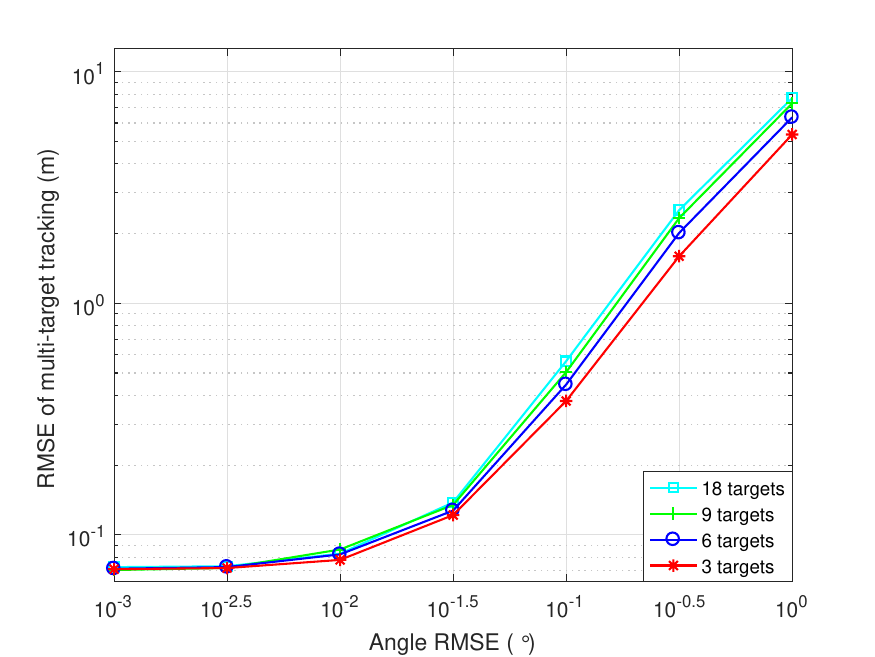}
 \caption{Multi-target tracking RMSEs versus angle errors under different number of targets ($v=3$m/s).}
 \label{locRMSE_TargetInvariation_vs_AngleError_diffTarNum}
\end{figure}

In \figurename{ \ref{locRMSE_TargetInvariation_vs_AngleError_diffTarNum}}, we exhibit the multi-target tracking RMSEs versus angle estimation errors under different target numbers.
It is obvious that as the increase of the angle RMSE, the multi-target tracking RMSEs also gets worser.
Besides, when the angle RMSE is larger, the increase of tracking RMSEs grows faster.
Moreover, the more the number of targets, the higher the tracking RMSE.
This can be explained as follows.
When the number of targets increases within a certain observation area, the distance between different targets is closer, resulting in a higher possibility of mutual influence between the observed angular information of different targets.
Besides, it is worth noting that there is no significant performance loss as the number of targets increases, which indicates the effectiveness and robustness of the proposed scheme.
This is because that when the number of targets is larger, the influence between the measurements of each target gradually converges, resulting in the convergence of tracking error.

%\textcolor[rgb]{0.00,0.00,1.00}{Besides, it is worth noting that the multi-target tracking RMSE increases slowly with the increase of the target amount when the number of targets is large.
%This is because when the number of targets is larger, the influence between the observations of each target gradually converges, resulting in the convergence of tracking error.}

\subsection{Numerical Results with Target Number Varying}

\begin{figure}[htbp]
 \centering
 \includegraphics[width=90mm]{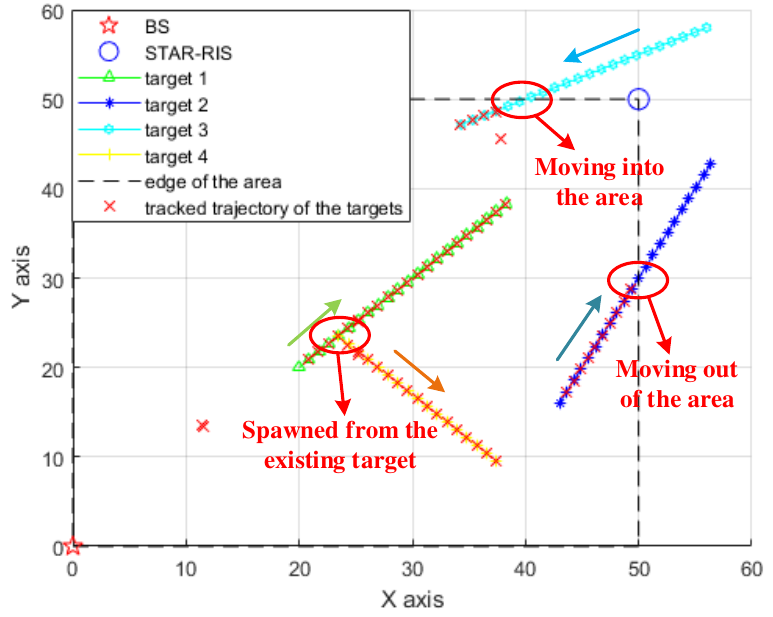}
 \caption{Vertical view for the illustration of multi-target tracking with target spawning and leaving.}
 \label{scenario_illustration_tarVariant}
\end{figure}

We then evaluate the proposed scheme under more complicated scenario (see \figurename{ \ref{scenario_illustration_tarVariant}}), where the angle RMSE is set as $0.001^\circ$,
and the number of the targets varies during the considered area and time duration.
Initially, only 2 targets are located in the observation area.
At the 5-th period, a new target is spawned from an existing target (see the yellow curve).
Besides, at the 12-th period, one target moves out of the observation area (see the blue curve).
Moreover, at 17-th period, a new target moves into the observation area (see the cyan curve).
We can observe that the proposed scheme can still successfully track the dynamic multi-targets even in such complicated scenario.

\begin{figure}[htbp]
 \centering
 \includegraphics[width=95mm]{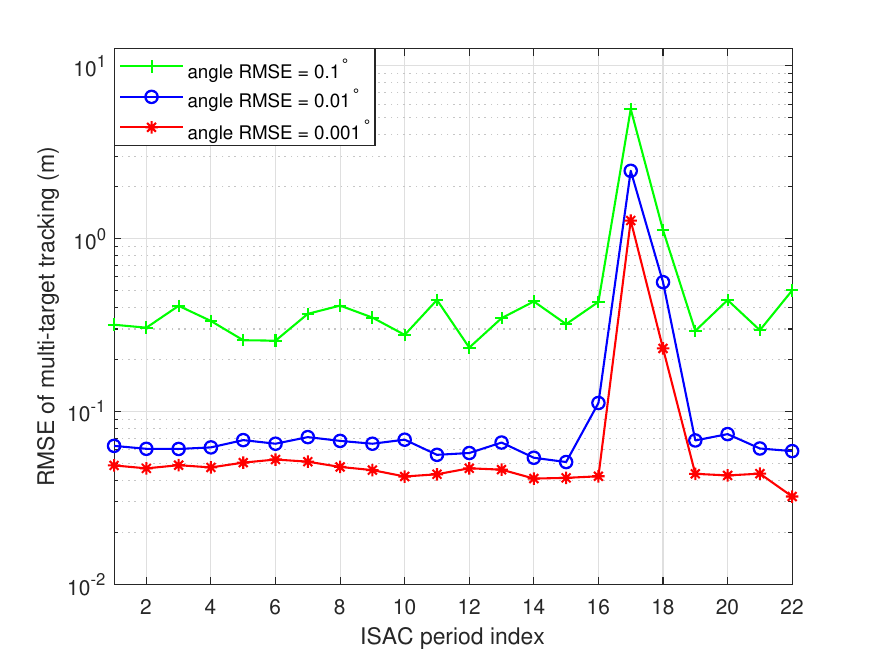}
 \caption{Multi-target tracking RMSEs versus ISAC period indexes under different level of angle errors ($v=3$m/s).}
 \label{locRMSE_TargetVariation_vs_Timestep_differentAngleError}
\end{figure}

\begin{figure}[htbp]
 \centering
 \includegraphics[width=95mm]{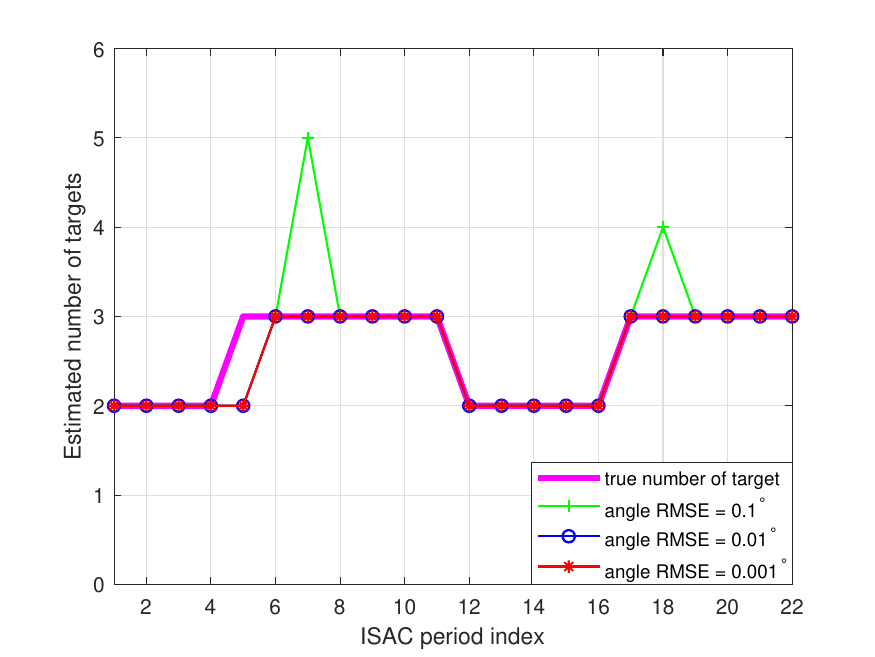}
 \caption{Estimated number of targets versus ISAC period indexes under different level of angle errors ($v=3$m/s).}
 \label{estTarNum_TargetVariation_vs_Timestep_differentAngleError}
\end{figure}

\figurename{ \ref{locRMSE_TargetVariation_vs_Timestep_differentAngleError}} and \figurename{ \ref{estTarNum_TargetVariation_vs_Timestep_differentAngleError}} respectively show the RMSE of multi-target tracking and the estimated number of targets versus the ISAC period index under different level of angle estimation errors.
Similar to the results in \figurename{ \ref{locRMSE_TargetInvariation_vs_Timestep_differentAngleError}}, the tracking RMSEs decrease with the increase of angle estimation accuracy.
Besides, although the number of targets has changed three times within the considered area and time duration as shown in Fig. 10, the tracking RMSEs still have negligible fluctuations at the 5-th and the 12-th period.
However, at the 17-th period, it is obvious that the tracking RMSEs ascend to a very high level and then descend back to their own average level.
Such observations can be explained as follows.
At the 5-th period, since the spawned target is related to an existing target, we can apply the a prior information to track the spawned target fastly.
In addition, at the 12-th period, there is no measurement related to the leaving target, and thus the tracking performance will not be influenced.
However, since we do not have any prior information about the new target entering at the 17-th period, an amount of observations are required to identify and track this new target.
Fortunately, our proposed scheme can track back this new target in just a few ISAC periods.

\begin{figure}[htbp]
 \centering
 \includegraphics[width=95mm]{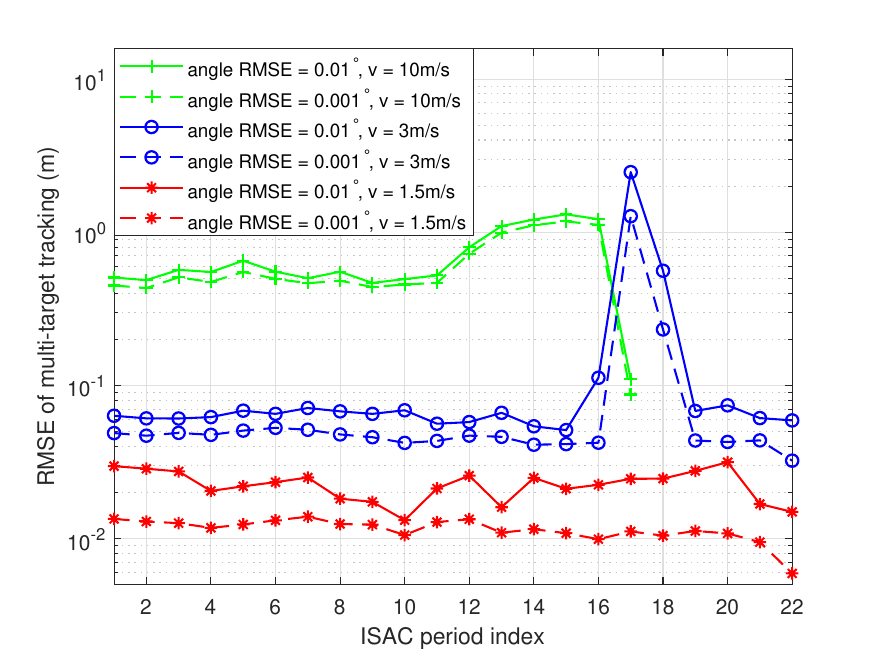}
 \caption{Multi-target tracking RMSEs versus ISAC period indexes under different level of velocities.}
 \label{locRMSE_TargetVariation_vs_Timestep_differentVelocity}
\end{figure}

In \figurename{ \ref{locRMSE_TargetVariation_vs_Timestep_differentVelocity}}, the RMSEs of multi-target tracking under different angle RMSEs and different target velocities are presented.
We only show the performance with angle RMSE $= 0.01^{\circ}$ and $0.001^{\circ}$ here since the tracking performance is stable when the angle error is less than $0.01^{\circ}$ as observed in \figurename{ \ref{locRMSE_TargetInvariation_vs_AngleError_diffTarNum}}.
From \figurename{ \ref{locRMSE_TargetVariation_vs_Timestep_differentVelocity}}, it can be observed that the higher the velocity, the worse the tracking performance.
Besides, with higher velocity, the performance under different angle errors tends to be closer.
This can be explained as follows.
When the velocity is high, the correlation between the observations at neighboring ISAC periods gets lower, which will cause the deterioration of the tracking performance.
Under such circumstance, the influence of angle estimation accuracy becomes lower than that of the target velocities.
In addition, it is worth mentioning that the green curves (i.e., velocity is 10 m/s) are terminated around 17-th ISAC period index.
This is due to the fact that the targets are all moving out of the considered area with such high velocity after a certain period of time.

\begin{figure}[htbp]
 \centering
 \includegraphics[width=95mm]{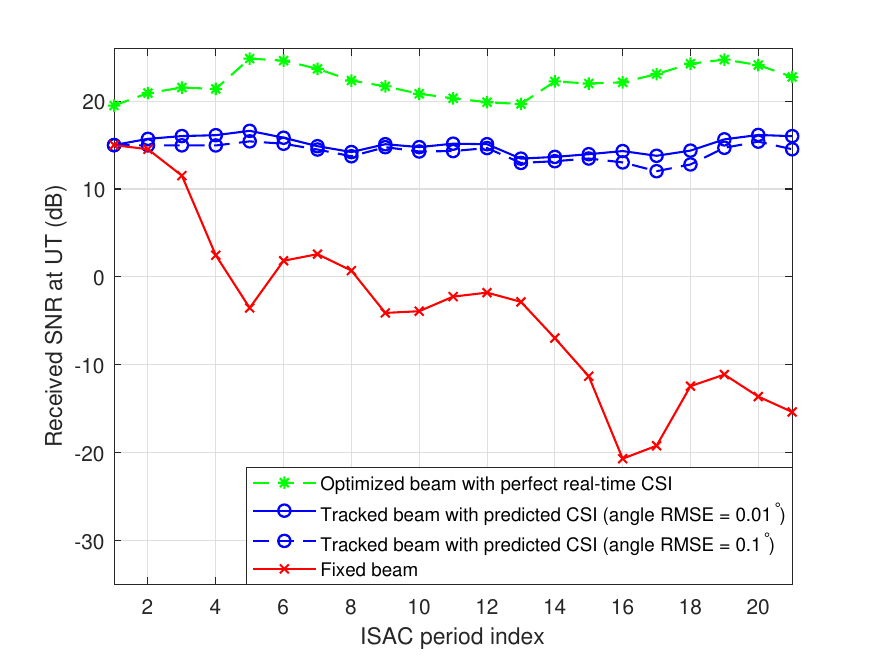}
 \caption{The received SNR at UT versus ISAC period indexes under different beam settings.}
 \label{Received_SNR_beam_prediction}
\end{figure}

\figurename{ \ref{Received_SNR_beam_prediction}} illustrates the beam prediction performance, where the received SNR at the indoor UT are compared for different beam settings.
%Note that the noise power is set to be 15 dB lower than the power of the UT received signal of the first sensing period.
Note that the fixed beam means that the BS only utilizes the designed beam of the first ISAC period without tracking.
It is obvious that the received SNR by using optimized beam with perfect real-time channel state information (CSI) is the highest, while that with fixed beam is the worst.
Moreover, it can be checked that the received SNR with fixed beam is very unstable, and becomes increasingly unacceptable with the sensing period growing.
By contrast, the received SNRs with tracked beam are very stable, and close to that with the optimized beams.
In addition, when the angle RMSE becomes lower, the received SNR at the UT with the tracked beam gets higher.
Note that our proposed scheme is to track the beams with predicted CSI, while the acquisition of the optimized beams requires the real-time CSI, which is very difficult to realize.
%\textcolor[rgb]{1.00,0.00,0.00}{PX: beam pattern: [precise pattern] vs [pattern not tracking in later time step] vs [pattern target tracking]
%[X: angles], [Y: ]. [Label: different situations.]}

\section{Conclusion}
In this paper, we proposed a multi-target tracking scheme for the STAR-RIS aided downlink ISAC system, where the STAR-RIS is utilized for simultaneous communication service of the indoor UT and sensing and tracking of the outdoor dynamic scatterers.
Firstly, we presented the scenario configuration, and introduced the state transition model of the outdoor dynamic scatterers.
We then developed the corresponding transmission frame structure for the downlink ISAC scheme.
Moreover, the channel models of the BS-RIS link and the RIS-UT link were constructed, followed by the formulation of received signal models respectively at the active STAR-RIS elements and the indoor UT.
With the knowledge related to the channel parameters of the BS-RIS link, we identified the dynamic scattering paths from the entire channel, and classified the dynamic scatterers with respect to their RCSs.
Furthermore, we proposed to track a varying number of scatterers on the STAR-RIS via the GM-PHD filter.
To further enhance communication performance of the indoor UT, we developed a beam prediction strategy for the BS precoder and STAR-RIS refraction phase shift vector.
Besides, we proposed a target mismatch detection and path collision prediction mechanism to reduce the training overhead and improve the transmission performance.
Finally, simulation results were provided to validate the feasibility and effectiveness of the proposed tracking scheme for STAR-RIS aided ISAC system.

\balance

%\balance
%\bibliographystyle{IEEEtran}
%\bibliography{./bibtex/IEEEabrv,./bibtex/ref}

%\balance
%\bibliographystyle{IEEEtran}
%\bibliography{./bibtex/IEEEabrv,./bibtex/ref}
\end{document}